\documentclass[english]{revtex4-1}
\usepackage[T1]{fontenc}
\usepackage[latin9]{inputenc}
\usepackage{geometry}
\usepackage{units}
\usepackage{amsmath}
\usepackage{graphicx}
\usepackage{setspace}
\usepackage{esint}
\usepackage{babel}

\begin{document}
\title{Macroscopic Fluctuations Emerge in Balanced Networks with Incomplete
Recurrent Alignment}
\author{Itamar D. Landau\footnote{corresponding author: itamar.landau@mail.huji.ac.il, current affiliation: Stanford University}}
\affiliation{Center for Brain Sciences, Hebrew University of Jerusalem, Israel}

\author{Haim Sompolinsky}

\affiliation{Center for Brain Sciences, Hebrew University of Jerusalem, Israel}
\affiliation{Racah Institute of Physics, Hebrew University of Jerusalem, Israel}
\affiliation{Center for Brain Sciences, Harvard University, Cambridge MA}
\date{March 11, 2021}
\maketitle
\global\long\def\xnia#1#2#3{\left(#1_{1},\cdots,#2_{#3}\right)}%
\global\long\def\xvnia#1#2{\left(\begin{array}{c}
#1\\
\vdots\\
#2
\end{array}\right)}%
\global\long\def\at#1#2#3{\left.#1\right|_{#2}^{#3}}%
 
\global\long\def\lag{\mathcal{L}}%
\global\long\def\f{\varphi}%
\global\long\def\t{\theta}%
\global\long\def\xep{\varepsilon}%
\global\long\def\hil{\mathcal{H}}%
 
\global\long\def\xp{\mathbb{P}}%
\global\long\def\xn{\mathbb{N}}%
\global\long\def\xz{\mathbb{Z}}%
\global\long\def\xrn#1{\mathbb{R}^{#1}}%
\global\long\def\xq{\mathbb{Q}}%
\global\long\def\xc{\mathbb{C}}%
\global\long\def\re{\mathrm{Re}}%
\global\long\def\im{\mathrm{Re}}%
\global\long\def\fourier{\mathcal{F}}%

\global\long\def\xd{\mathcal{D}}%
\global\long\def\parfunc{\mathcal{Z}}%
\global\long\def\xnorm{\mathcal{N}}%
\global\long\def\xb{\mathcal{B}}%
\global\long\def\riemann{\mathcal{R}}%
~
\global\long\def\xs{\mathcal{S}}%
\global\long\def\xP{\mathcal{P}}%
\global\long\def\xC{\mathcal{C}}%
\global\long\def\order{\mathcal{O}}%
\global\long\def\xQ{\mathcal{Q}}%
 
\global\long\def\xk{\mathcal{K}}%
\global\long\def\xI{\mathcal{I}}%

\global\long\def\r{\vec{r} }%
\global\long\def\case#1{\left\{  #1\right.}%

\global\long\def\pdr#1#2{\dfrac{\partial#1}{\partial#2}}%
\global\long\def\pddr#1#2{\dfrac{\partial^{2}#1}{\partial#2^{2}}}%
\global\long\def\pdrn#1#2{\dfrac{\partial^{n}#1}{\partial#2^{n}}}%
\global\long\def\pa{\partial}%
\global\long\def\ppx#1#2#3{\frac{\partial^{2}#1}{\partial#2\partial#3}}%

\global\long\def\dr#1#2{\dfrac{\mathrm{d}#1}{\mathrm{d}#2}}%
\global\long\def\ddr#1#2{\dfrac{\mathrm{d^{2}}#1}{\mathrm{d}#2^{2}}}%
\global\long\def\drn#1#2{\dfrac{\mathrm{d}^{n}#1}{\mathrm{d}#2^{n}}}%
\global\long\def\Dr#1#2{\frac{D#1}{D#2}}%
\global\long\def\drdt#1{\dfrac{\mathrm{d}#1}{\mathrm{d}t}}%
\global\long\def\drdx#1{\dfrac{\mathrm{d}#1}{\mathrm{d}x}}%

\global\long\def\mod{\text{mod }}%
 
\global\long\def\sgn{\mathrm{sgn}}%

\global\long\def\e{\mathrm{\mathbb{E}}}%
\global\long\def\var{\mathrm{Var}}%
\global\long\def\pcon#1#2{\mathbb{P}\left(#1|#2\right)}%
\global\long\def\cov{{\rm Cov}}%
\global\long\def\erf#1{\mathrm{erf}\left(#1\right)}%

\global\long\def\norm#1{\left\Vert #1\right\Vert }%
\global\long\def\argmin#1#2{\underset{#1}{\mathrm{argmin}}#2}%
\global\long\def\const{\mathrm{const}}%
\global\long\def\argmax#1#2{\underset{#1}{\mathrm{argmax}}#2}%
\global\long\def\trace#1{\underset{#1}{\mathrm{Tr}}}%

\global\long\def\rm#1{\mathrm{#1}}%
\global\long\def\reac#1#2{\overset{#1}{\underset{#2}{\leftrightharpoons}}}%
\global\long\def\adj#1{\mathrm{adj}\left(#1\right)}%
\global\long\def\span#1{\mathrm{span}\left(#1\right)}%
\global\long\def\up{\uparrow}%
\global\long\def\do{\downarrow}%
\global\long\def\ulim#1{\overset{#1}{\longrightarrow}}%
\global\long\def\xlim#1#2{\lim_{#1\rightarrow#2}}%
\global\long\def\block#1{\left|\;\begin{matrix}\hline #1\\\hline \end{matrix}\;\right|}%
\global\long\def\sect{\mathsection}%
\global\long\def\fder#1#2{\frac{\delta#1}{\delta#2}}%
\global\long\def\bb#1{\boldsymbol{#1}}%
\global\long\def\ex#1{\mathrm{\mathrm{e}}^{#1}}%
\global\long\def\fl#1{\left\lfloor #1\right\rfloor }%

\section*{Abstract}

Networks of strongly-coupled neurons with random connectivity exhibit
chaotic, asynchronous fluctuations. In a previous study, we showed
that when endowed with an additional low-rank connectivity consisting
of the outer product of orthogonal vectors, these networks generate
large-scale coherent fluctuations. Although a striking phenomenon,
that result depended on a fine-tuned choice of low-rank structure.
Here we extend that result by generalizing the theory of excitation-inhibition
balance to networks with arbitrary low-rank structure and show that
low-dimensional variability emerges intrinsically through what we
call ``incomplete recurrent alignment''. We say that a low-rank
connectivity structure exhibits incomplete alignment if its row-space
is not contained in its column-space. In the generic setting of incomplete
alignment, recurrent connectivity can be decomposed into a ``subspace-recurrent''
component and an ``effective-feedforward'' component. We show that
high-dimensional, microscopic fluctuations are propagated via the
effective-feedforward component to a low-dimensional subspace where
they are dynamically balanced by macroscopic fluctuations. We present
biologically plausible examples from excitation-inhibition networks
and networks with heterogeneous degree distributions. Finally, we
define the alignment matrix as the overlap between left- and right-singular
vectors of the structured connectivity, and show that the singular
values of the alignment matrix determine the amplitude of low-dimensional
variability, while its singular vectors determine the structure. Our
work shows how low-dimensional fluctuations can emerge generically
in strongly-coupled networks with low-rank structure. Furthermore,
by generalizing excitation-inhibition balance to arbitrary low-rank
structure our work may find relevance in any setting with strongly
interacting units, whether in biological, social, or technological
networks.

\part*{\protect\pagebreak Introduction}

The dynamic balance of excitation (E) and inhibition (I) is a paradigmatic
theory for describing the activity of neocortical networks \citep{VanVreeswijk1996,VanVreeswijk1998,Brunel2000,Renart2010b}.
The theory describes how recurrent interactions generate asynchronous
irregular firing activity which is typically observed in the cortex
across many species, especially in awake, active states of behavior
\citep{Softky1993,Ecker2010,Ecker2014,Cohen2011,Doiron2016}. The
E-I balance network model is driven by strong feed-forward excitation,
and strong recurrent connectivity, dominated by inhibition, is necessary
in order to balance that input. The result is that the average activity
of E and I populations dynamically balance the mean synaptic input,
enabling fluctuations to propagate asynchronously. The resulting networks
can account for various empirical observations of cortical activity,
including not only irregular firing but also low pairwise correlations
\citep{Renart2010b} and broad firing-rate distributions \citep{Roxin2011a}.
Furthermore, balanced networks perform fast-tracking of external input
which can underly predictive coding \citep{Kadmon2020}, and they
are capable of amplifying input feature selectivity \citep{Hansel2012,Pehlevan2014}
and generating stable patterns of activity for associative memory
\citep{Vreeswijk2005,Roudi2007,Mongillo2018}. \\

Many qualitative aspects of the dynamics of excitation-inhibition
balance can be understood by studying simpler firing-rate models \citep{Wilson1972}.
Firing rate models of randomly connected excitatory and inhibitory
populations with strong interactions and strong feed-forward input
exhibit a dynamic cancelation of the mean input at the population
level \citep{Harish2015,Kadmon2015}, similarly to the spiking models.
The resulting state is chaotic and asynchronous, due to asymmetric
random connections\citep{Sompolinsky1988}.\\

It had been a longstanding question whether such randomly connected
networks could intrinsically generate large-scale coherent fluctuations.
We previously showed that when randomly connected networks are endowed
with additional low-rank connectivity of a particular structure, fluctuations
emerge that are shared coherently across the entire network \citep{Landau2018}.
Specifically, connectivity structure consisting of outer-products
of orthogonal pairs embed a purely feedforward structure into the
recurrent network such that fluctuations along the row-space are propagated
to the column-space, yielding shared variability in the column-space,
without generating feedback that would either supress fluctuations
or drive saturation. The same qualitative phenomenon was studied also
by Darshan et al \citep{Darshan2020} and Hayakawa and Fukai \citet{Hayakawa2020}.\\

In the current work we study a broader framework of what we refer
to as ``incomplete recurrent alignment,'' of which an orthogonal
outer-product is one limiting case. In order to develop our theory
of incomplete recurrent alignment, we generalize the theory of excitation-inhibition
balance to networks with arbitrary low-rank structure. A number of
recent studies have explored the dynamics of networks with low-rank
structured connectivity in addition to a random component of connectivity,
yet these have all focused on weakly-coupled low-rank structure \citep{Rivkind2017,Mastrogiuseppe2017a}.
We show that embuing random networks with strong low-rank connectivity
yields a natural generalization of excitation-inhibition balance:
a regime of dynamic balance in which strong input to a low-dimensional
subspace is canceled without fine-tuning, and fluctuations can emerge
in the orthogonal complement.\\

Using our generalized dynamic balance formalism, we show that incomplete
recurrent alignment genereates macroscopic fluctuations. We say that
a strongly-coupled recurrent network has incomplete alignment if the
row-space of its structured connectivity is not entirely contained
in its column-space. We decompose the structrued connectivity into
a ``subspace recurrent'' component, through which macroscopic activity
in the column-space is able to dynamically balance its input, and
an ``effective-feedforward'' component, through which microscopic
fluctuations in the orthogonal subspace serve as a source of fluctuating
input to the macroscopic dynamics in the column-space.\\

In the general case of incomplete alignment, this fluctuating source
from the orthogonal subspace is dynamically balanced by macroscopic
fluctuations in the column-space, which we refer to as the ``balance
subspace''. This balancing of fluctuations is analogous to the way
a balanced network cancels shared fluctuations received from external
sources \citep{Renart2010a}, except that here the source of the fluctuating
input is recurrent. The larger the extent of misalignment, the larger
the macroscopic fluctuations that arrise in order to achieve balance.
Our theory yields a second-order balance equation for intrinsically
generated macroscopic fluctuations, and we show how the macroscopic
correlation structure is fully determined by the overlap matrix between
left- and right-singular vectors of the structured connectivity, while
the time-course of fluctuations is inherited from the time-course
of microscopic fluctuations. \\

In Section \ref{sec:Model} we introduce the model. In Section \ref{sec:Balance-Subspace-Decomposition}
we present the decomposition into macroscopic order parameters that
reside in the low-dimensional ``balance subspace'' on the one hand,
and microscopic degrees of freedom in the orthogonal subpsace on the
other. We show that strong low-rank connectivity yields dynamic balance
-- a linear equation for the macroscopic firing rates in the balance
subspace, and microscopic chaotic fluctuations in the orthogonal subspace.
In Section \ref{sec:Incomplete-Align Fluctuations} we show that incomplete
alignment of the low-rank connectivity projects the microscopic fluctuations
into the balance subspace yielding amplified shared variability. We
derive expressions for the amplitude, spatial structure and timescale
of this variability. Finally, in Section \ref{sec:Biologically-Relevant-Examples}
we study two concrete examples of biologically relevant incomplete
alignment, excitatory-inhibitory networks with degenerate synaptic
weight parameters (as previously studied in \citep{Helias2014}),
and networks with heterogeneous out-degrees.

\part*{Results}

\section{Model\label{sec:Model}}

We study a network of $N$ firing-rate neurons with a connectivity
matrix consisting of structured and random components. The structured
component, $\mathbf{M}$, is given by a rank $D$ matrix, where $D$
is finite in the large $N$ limit, and the random component $\mathbf{J}$
has i.i.d. components assumed for simplicity to be Gaussian. Individual
elements of both components of the connectivity scale as $\frac{1}{\sqrt{N}}$.
Explicitly, we write the structured connectivity matrix in reduced
singular value decomposition (SVD) form:
\begin{equation}
\mathbf{M}=\frac{1}{\sqrt{N}}\mathbf{U}\bb{\Sigma}\mathbf{V}^{T}\label{eq:M}
\end{equation}
where both $\mathbf{U}$ and $\mathbf{V}$ are $N$-by-$D$ matrices
with $O\left(1\right)$ components and orthogonal columns of norm
$\sqrt{N}$, i.e. $\mathbf{U}^{T}\mathbf{U}=\mathbf{V}^{T}\mathbf{V}=N\mathbf{I}_{DxD}$
(notice that we adopt here a scaling of U and V that differs from
the standard SVD convention). $\bb{\Sigma}$ is a diagonal $D$-by-$D$
matrix with $O\left(1\right)$ positive elements, $\sigma_{k}$.\\

We define the random component, $\mathbf{J}$, whose elements are
sampled i.i.d. from $J_{ij}\sim\xnorm\left(0,\frac{g^{2}}{N}\right)$
with a ``gain'' parameter $g=O(1)$. For simplicity, we assume that
the elements of $\mathbf{J}$ are drawn independently of the structure
of $\mathbf{U}$ and $\mathbf{V}$, i.e., we assume that $\frac{1}{N}\mathbf{V}^{T}\mathbf{J}^{k}\mathbf{U}\approx0$
for all $k\ge1$\citet{Schuessler2020}.\\

The dynamics of the inputs (Fig 1A) are driven by strong external
drive, $\sqrt{N}\mathbf{f},$with $f_{i}\sim O\left(1\right)$, and
given by:
\begin{equation}
\drdt{\mathbf{h}}=-\mathbf{h}+\left(\mathbf{M}+\mathbf{J}\right)\mathbf{r}+\sqrt{N}\mathbf{f}\label{eq:Full Dynamics}
\end{equation}
where the firing rates of individual neurons are given by $r_{i}=\text{\ensuremath{\phi\left(h_{i}\right)}}$,
an instantaneous, sigmoidal non-linearity of the inputs. Unless otherwise
mentioned, simulations and numerical calculations in this paper use
$\phi\left(h)\right)=\tanh\left(h)\right)$, which can be thought
of as the change in firing rate relative to some baseline. \\

As displayed in Figure 1A, the network structure can be understood
schematically: $\mathbf{v}_{k}$, i.e. the $k$th right-singular vector
of $\mathbf{M}$, performs a read-out of the network activity; this
readout scaled by $\sigma_{k}$ and fed back into the network along
the corresponding left-singular vector $\mathbf{u}_{k}$ (Fig 1A).\\

\begin{figure}
\includegraphics{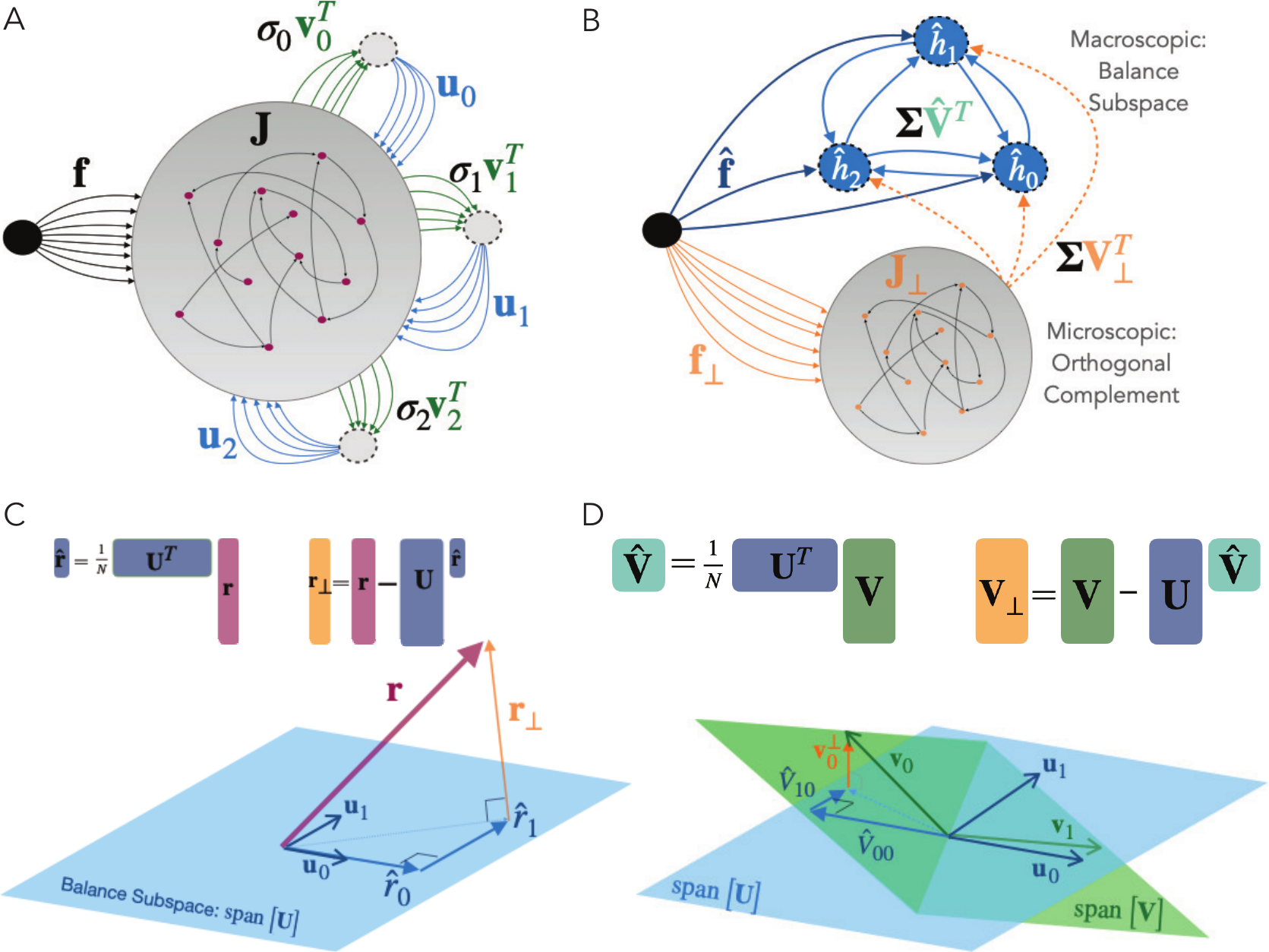}\caption{\textbf{Model Dynamics and Schematic.} \textbf{(A) }Schematic representation
of network structure with external drive vector, $\mathbf{f}$, random
recurrent connectivity, $\mathbf{J}$, and structured recurrent connectivity
represented as a read-out along each right-singular vector, $\mathbf{v}_{k}$,
scaled by $\sigma_{k}$, and fedback to the network along left-singular
vector, $\mathbf{u}_{k}$. \textbf{(B) }Balance subspace decomposition.
Schematic representing the decomposition into balance subspace and
the orthogonal complement. The subspace recurrence within the balance
subspace is given by $\protect\bb{\Sigma}\hat{\mathbf{V}}^{T}$. The
effective-feedforward connectivity from the orthogonal complement
to the balance subspace is given by $\protect\bb{\Sigma}\mathbf{V}_{\perp}^{T}$,
which is $0$ in the fully aligned case. \textbf{(C) }Balance subspace
decomposition for population vector of firing rates, $\mathbf{r}$.
Top: Equations\textbf{ }defining $\hat{\mathbf{r}}$ and $\mathbf{r}_{\perp}$.
Bottom: Geometric visualization of the balance subspace decomposition.The
projection into the balance subspace is given by the compontents,
$\hat{r}_{k}$, along each left-singular vector, $\mathbf{u}_{k}$.
The projection onto the orthgonal complement is $\mathbf{r}_{\perp}$.
The same definitions apply for\textbf{ }the dynamical variables, $\mathbf{h}$,
and the external drive, $\mathbf{f}$. \textbf{(D) }Balance subspace
decomposition of the connectivity structure, defining the alignment
matrix, $\hat{\mathbf{V}}$. Top: Equations defining $\hat{\mathbf{V}}$
and $\mathbf{V}_{\perp}$. The same definitions as in (C) are applied
to a matrix of $D$ column vectors, each of length $N$. Bottom: Geometrix
visualization of incomplete alignment. The span of the columns of
$\mathbf{V}$ may generally not be constrained to the balance subspace.
The right-singular vector, $\mathbf{v}_{0}$, is projected onto the
balance subspace where its component along each left-singular vector,
$\mathbf{u}_{0}$ and $\mathbf{u}_{1}$, define the matrix elements
$\hat{V}_{00}$ and $\hat{V_{10}}$, respectively. The projection
of $\mathbf{v}_{0}$ onto the orthogonal complement defines, $\mathbf{v}_{0}^{\perp}$,
the first column of $\mathbf{V}_{\perp}$. Similarly for all right-singular
vectors, not shown. If the connectivty structure is ``fully aligned'',
then the span of $\mathbf{V}$ is identical to the span of $\mathbf{U}$
and the right-singular vectors are obtained from the left-singular
vectors by rotation and reflection, such that the alignment matrix,
$\hat{\mathbf{V}}$, is orthonormal.}
\end{figure}

\section{Balance Subspace Decomposition and the Alignment Matrix\label{sec:Balance-Subspace-Decomposition}}

The span of the left-singular vectors, $\mathbf{u}_{k}$, i.e the
column-space of the structured connectivity $\mathbf{M}$, defines
a distinct subspace. Both the structured and random components of
connectivity have strong single synapses, i.e. they are $O\left(\frac{1}{\sqrt{N}}\right)$.
However, we observe that the SVD of the random component will yield
singular values of $O\left(\frac{1}{N}\right)$, such that any vector
of $O\left(1\right)$ firing rates will yield only $O\left(1\right)$
input via the random component. On the other hand, the structured
component of connectivity can yield $O\left(\sqrt{N}\right)$ input
along the columns of $\mathbf{U}$. Therefore, the subspace spanned
by the columns of $\mathbf{U}$ can receive strong recurrent inputs
that will drive saturation unless they are balanced, and it will define
the macroscopic order parameters of our system.\\

We call the column-space of the structured connectivity the ``balance
subspace''. We define the decomposition into the balance subspace
and its orthogonal complement: for any $N$-dimensional vector or
matrix with $N$-dimensional column-space, $\mathbf{X}$, we write
$\mathbf{X}=\mathbf{X}_{\parallel}+\mathbf{X}_{\perp}$, where $\mathbf{X}_{\parallel}=\frac{1}{N}\mathbf{U}\mathbf{U}^{T}\mathbf{X}$
is the projection in the balance subspace, and $\mathbf{X}_{\perp}=\left(\mathbf{I}-\frac{1}{N}\mathbf{U}\mathbf{U}^{T}\right)\mathbf{X}$
is the projection in the orthogonal subspace.\\

We study the components in the balance subspace relative to the columns
of $\mathbf{U}$, that is we write $\mathbf{X}_{\parallel}=\mathbf{U}\hat{\mathbf{X}}$
with 
\begin{equation}
\hat{\mathbf{X}}\equiv\frac{1}{N}\mathbf{U}^{T}\mathbf{X}\label{eq:Bal Subspace Projection}
\end{equation}

Figure 1C displays the geometry of this decomposition for the population
firing rate vector, $\mathbf{r}=\mathbf{U}\hat{\mathbf{r}}+\mathbf{r}_{\perp}$.
We will similarly decompose the input currents, $\mathbf{h}$, into
$\mathbf{U}\hat{\mathbf{h}}+\mathbf{h}_{\perp}$.\\

Additionally, as will be motivated in the following section, we will
decompose the matrix of right-singular vectors, $\mathbf{V}=\mathbf{U}\hat{\mathbf{V}}+\mathbf{V}_{\perp}$,
introducing the ``alignment matrix'', $\hat{\mathbf{V}}$, between
the column-space and row-space of the structured connectivity, 
\begin{equation}
\hat{\mathbf{V}}=\frac{1}{N}\mathbf{U}^{T}\mathbf{V}
\end{equation}
The alignment matrix is the $D$-by-$D$ matrix consisting of the
overlap of each right-singular vector of the structured connectivity
along each left-singular vector. Figure 1D displays the geometry of
the two overlapping hyperplanes, given by the span of left- and right-singular
vectors \\

Importantly, if the right-singular vectors are all contained in the
balance subspace, i.e. if the row-space is entirely contained in the
column-space, thenthe orthogonal complement is null ($\mathbf{V}_{\perp}=\mathbf{0}$).
We refer to this condition as ``full recurrent alignment''. Full
recurrent alignment is obtained if an only if $\hat{\mathbf{V}}^{T}\hat{\mathbf{V}}=\mathbf{I}_{DxD}$,
i.e. if the alignment matrix is orthonormal.\\

We will assume the external drive is aligned with the balance subpsace
so that $\mathbf{f}=\mathbf{U}\hat{\mathbf{f}}$.

\subsection{Balance Subspace Dynamics}

We decompose the dynamics \ref{eq:Full Dynamics}, and first study
the the dynamics in the balance subspace, which are the macroscopic
order parameters of the network. To do so we apply the above decomposition
(Eqn\ref{eq:Bal Subspace Projection}) on the structured connectivity,
$\mathbf{M}$, Eq. \ref{eq:M}, in order to write $\hat{\mathbf{M}}=\frac{1}{\sqrt{N}}\mathbf{\bb{\Sigma}}\mathbf{V}^{T}$.
We note that by construction, $\mathbf{M}_{\perp}=0$, since $\mathbf{M}$
projects entirely into the balance subspace.\\

We use the alignment matrix, $\hat{\mathbf{V}}$, and the orthogonal
complement\textbf{ ,$\mathbf{V}_{\perp}$}, to further decompose $\mathbf{M}$,
giving $\hat{\mathbf{M}}=\frac{1}{\sqrt{N}}\bb{\Sigma}\hat{\mathbf{V}}^{T}\mathbf{U}^{T}+\frac{1}{\sqrt{N}}\bb{\Sigma}\mathbf{V}_{\perp}^{T}$.
The first term drive the ``subspace-recurrent'' component of the
input, which is due to activity within the balance subspace, while
the second term drives the ``effective-feedforward'' input, which
is due to activity in the orthogonal complement.\\

Given population firing rates, $\mathbf{r}=\mathbf{U}\mathbf{\hat{r}}+\mathbf{r}_{\perp}$,
the subspace-recurrent input is $\sqrt{N}\bb{\Sigma}\hat{\mathbf{V}}^{T}\hat{\mathbf{r}}$
and the effective-feedforward input is $\frac{1}{\sqrt{N}}\bb{\Sigma}\mathbf{V}_{\perp}^{T}\mathbf{r}_{\perp}$.
Projecting the dynamics (Eqn \ref{eq:Full Dynamics}) onto the balance
subspace yields the macroscopic dynamcis:

\begin{equation}
\drdt{\hat{\mathbf{h}}}=-\hat{\mathbf{h}}+\sqrt{N}\left(\bb{\Sigma}\hat{\mathbf{V}}^{T}\hat{\mathbf{r}}+\hat{\mathbf{f}}\right)+\frac{1}{\sqrt{N}}\bb{\Sigma}\mathbf{V}_{\perp}^{T}\mathbf{r}_{\perp}+O\left(\frac{1}{\sqrt{N}}\right)\label{eq:Balance Subspace Dynamics}
\end{equation}

where the $O\left(\frac{1}{\sqrt{N}}\right)$ term is the contribution
from the random connectivity, $\mathbf{J}$, onto the balance subspace,
which we have ignored. Note that the effective-feedforward contribution
via $\mathbf{V}_{\perp}$ is of $O\left(1\right)$. \\
\\
These macroscopic dynamics admit a balanced fixed point governed by
$D$ linear equations:
\begin{equation}
\bb{\Sigma}\hat{\mathbf{V}}^{T}\hat{\mathbf{r}}^{*}+\hat{\mathbf{f}}=O\left(\frac{1}{\sqrt{N}}\right)\label{eq:Linear Bal Eqns}
\end{equation}

Note that the effective-feedforward input does not contribute to the
balance fixed point up to leading order. It will nevertheless have
an impact on the macroscopic fluctuations around the fixed point as
we shall see below.\emph{}\\

The linear balance equations and their solution generalize the $2$-dimensional
E-I balance equations to arbitrary low-rank structure and emphasize
that they have a natural basis in the columns of $\mathbf{U}$, i.e.
the left singular vectors of the structured component of connectivity.
Moreover these balance equations are independent of the structure
of $\mathbf{U}$, depending only on the singular values and the alignment
matrix $\hat{\mathbf{V}}$.\\

Finally, we note that in general the balance equations may or or may
not have an obtainable solution (e.g., consistent with non-saturating
local rates). In the following we will assume that the local firing
activations and external inputs are normalized such that the balance
equations yield a feasible solution, see Appendix \ref{subsec:Appendix-A--Self-consistency}
and \ref{subsec:Appendix-B--MF in the Orthogonal}.\\

\subsection{Microscopic dynamics}

We now consider the dynamics in the orthogonal complement, by projecting
the full dynamics (Eqn \ref{eq:Full Dynamics}) via $\mathbf{h}_{\perp}=\left(\mathbf{I}-\frac{1}{N}\mathbf{U}\mathbf{U}^{T}\right)\mathbf{h}$
(recall that we assume that the external drive is contained in the
balance subspace):

\begin{equation}
\drdt{\mathbf{h}_{\perp}}=-\mathbf{h}_{\perp}+\mathbf{J}_{\perp}\mathbf{r}
\end{equation}

Due to the random connectivity, $\mathbf{J}_{\perp}$, These microscopic
dynamics can be described by a Gaussian dynamic mean-field theory
which we detail in the Appendix following \citet{Kadmon2015}. The
mean-field theory predicts that for sufficiently strong random connectivity
these equations will generate asynchronous chaotic dynamics. The order
parameter of the chaotic state is the mean single-neuron autocorrelation
function $C\left(\tau\right)\equiv\left\langle \delta r_{i}\left(t\right)\delta r_{i}\left(t+\tau\right)\right\rangle $.
The implicit differential equation governing $C(\tau)$ is derived
from the dynamic mean-field theory and given in Appendix \ref{subsec:Appendix-B--MF in the Orthogonal}.
Note that these microscopic dynamics and the resulting mean-field
theory depend on the macroscopic dynamics in the balance subspace,
$\hat{\mathbf{h}}$. Both the fixed point values of $\hat{\mathbf{h}}$
and the autocorrelation, $C\left(\tau\right)$, must be consistent
with the firing rates determined by the balance equations. We solve
the dynamic mean-field theory and find a good match between simulation
and theory (SI Fig 1).\\

As mentioned above, we assume that the balance firing rates given
by Eqn \ref{eq:Linear Bal Eqns} yield a non-saturating solution such
that sufficiently strong random connectivity will yield chaotic microscopic
dynamics (See Appendix \ref{subsec:Appendix-A--Self-consistency}).

\begin{figure}
\includegraphics{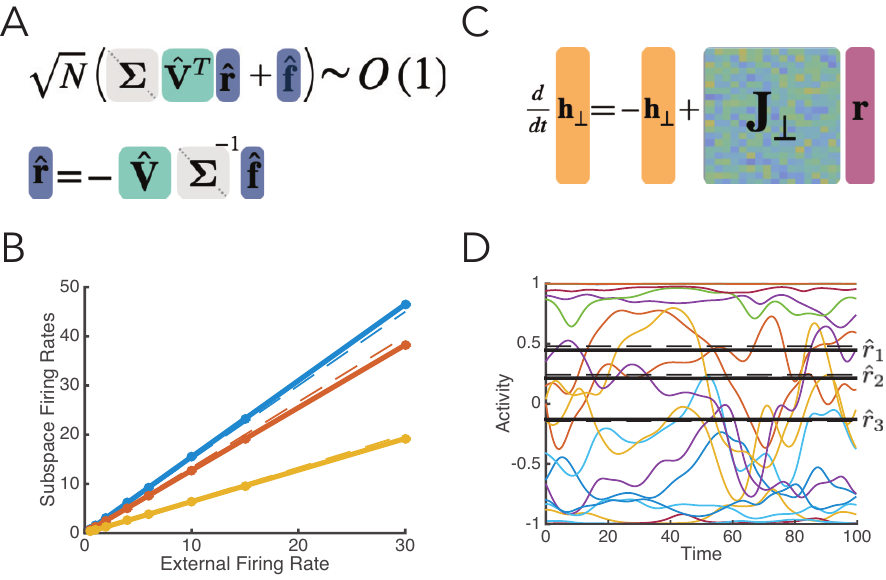}\caption{\textbf{Dynamic Balance in Networks with Low-Rank Structure. (A)}
Top: Balance requirement for the cancelation of potentially large
($\sqrt{N}$) input into the balance subspace. Bottom: Solution to
\textbf{$D$ }linear balance equations in the case where the column-space
and row-space are fully aligned\textbf{ (}alignment matrix, $\hat{\mathbf{V}}$,
is orthonormal)\textbf{. (B) }Balance subspace time-average firing
rates, $\left\langle \hat{r}_{k}\right\rangle $, vs overall scaling
factor of external firing rate, solid lines are the simulations which
match closely with the prediction from linear balance equations (A)
shown in dotted lines\textbf{ (C) }Dynamics of the microscopic degrees
of freedom in the orthogonal complement. \textbf{(D) }Sample activity
trace. Colored lines are individual neuron firing rates, $r_{i}\left(t\right)$,
displaying chaotic fluctuations. Black lines are the balance subspace
firing rates, $\hat{r}_{k}\left(t\right)$, with simulations in solid
lines and theory in dashed lines. ((B) employed a rectified-linear
non-linearity $\phi\left(h\right)\equiv\left\lfloor h\right\rfloor _{+}$)}

\end{figure}

\section{Incomplete Alignment Amplify Fluctuations\label{sec:Incomplete-Align Fluctuations}}

In this section we assume a network in the chaotic balanced state
described above. In this state, the macroscopic balance subspace firing
rates, $\hat{\mathbf{r}}$ (defined via \ref{eq:Bal Subspace Projection})
satisfy the balance equations (Eqn \ref{eq:Linear Bal Eqns}), and
are thus constant to leading order. \\

In this section we show that incomplete alignment amplifies macroscopic
fluctuations in the balance subspace, $\delta\hat{\mathbf{r}}\left(t\right)\equiv\hat{\mathbf{r}}\left(t\right)-\left\langle \hat{\mathbf{r}}\right\rangle $.
These are quantified by the $D$-by-$D$ matrix of covariance functions:
\begin{equation}
\hat{\mathbf{C}}\left(\tau\right)\equiv\left\langle \delta\hat{\mathbf{r}}\left(t\right)\delta\hat{\mathbf{r}}^{T}\left(t+\tau\right)\right\rangle 
\end{equation}
We will study the dynamics of the macroscopic, balance subspace fluctuations,
$\delta\hat{\mathbf{h}}\equiv\hat{\mathbf{h}}-\left\langle \hat{\mathbf{h}}\right\rangle $,
by subtracting the time-average from Eqn \ref{eq:Balance Subspace Dynamics}.
We first consider the case of full alignment, in which $\mathbf{V}_{\perp}=0$,
and there is no effective-feedforward input from the orthogonal subspace
to the balance subspace. The dynamics of $\delta\hat{\mathbf{h}}$
are then given by
\[
\drdt{\delta\hat{\mathbf{h}}}=-\delta\hat{\mathbf{h}}+\bb{\Sigma}\sqrt{N}\hat{\mathbf{V}}^{T}\delta\hat{\mathbf{r}}+O\left(\frac{1}{\sqrt{N}}\right)
\]
where the $O\left(\frac{1}{\sqrt{N}}\right)$ contribution is from
the random connectivity, and we remind the reader that the alignment
matrix, $\hat{\mathbf{V}}=\frac{1}{N}\mathbf{U}^{T}\mathbf{V}$, is
orthonormal in the case of full alignment.\\

In this case, we must have $\sqrt{N}\delta\hat{\mathbf{r}}\ll1$,
otherwise the input fluctuations driving $\delta\hat{\mathbf{h}}$
will be $O\left(1\right)$, and this will destabilize the balance
state. We verify this numerically and find that in fully algined networks
the total variance of temporal fluctuations, given by the trace of
$\hat{\mathbf{C}}\left(0\right)$, is $O\left(\frac{1}{N^{2}}\right)$
(as also discussed recently \citet{Kadmon2020}).\\

To probe the impact of misalignment, we simulate networks with varying
levels of alignment (see Appendix \ref{subsec:D1---Uniform Misalignment}
for details) and varying network size. We find that in misaligned
networks the macroscopic fluctuations are larger by an order of magnitude:
their total variance is $O\left(\frac{1}{N}\right)$ (Fig 3A) . Furthermore,
we find that incomplete alignment yields non-trivial covariance structure,
whereas in fully aligned networks $\hat{\mathbf{C}}$is essentially
structureless (Fig 3B).\\

To explain the emergence of macroscopic fluctuations, we return to
the dynamics of $\delta\hat{\mathbf{h}}$, in the general, $\mathbf{V}_{\perp}\neq0$,
case:
\begin{equation}
\drdt{\delta\hat{\mathbf{h}}}=-\delta\hat{\mathbf{h}}+\bb{\Sigma}\left(\sqrt{N}\hat{\mathbf{V}}^{T}\delta\hat{\mathbf{r}}+\hat{\bb{\eta}}\right)\label{eq:Fluctuation Dynamics}
\end{equation}

where, $\hat{\bb{\eta}}=\frac{1}{\sqrt{N}}\mathbf{V}_{\perp}^{T}\delta\mathbf{r}_{\perp}$,
are the fluctuations in the effective-feedforward input from the orthogonal
subspace to the balance subspace due to the misaligned connectvity.
If the typical elements of $\mathbf{V}_{\perp}$ are $O\left(1\right)$
then microscopic fluctuations in the orthogonal subspace are projected
to the balance subspace yielding input-fluctuations, $\hat{\bb{\eta}}$,
that are $O\left(1\right)$. Thus we expect incomplete alignment to
drive significant macroscopic input correlations. Indeed, as we show
for an example network in Figure 3C, the net input-fluctuations from
the orthogonal subspace onto the balance subspace is $O\left(1\right)$.\\

A stable balanced state requires that these $O\left(1\right)$ effective
feed-forward input-fluctuations from $\hat{\bb{\eta}}$ be canceled
to leading order by recurrent balance subspace fluctuations in $\delta\hat{\mathbf{r}}$,
maintaining $O\left(\frac{1}{\sqrt{N}}\right)$ fluctuations in $\delta\hat{\mathbf{h}}$.
This argument yields a fluctuation balance equation for the macroscopic
activity order parameters:
\begin{equation}
\sqrt{N}\hat{\mathbf{V}}^{T}\delta\hat{\mathbf{r}}+\hat{\bb{\eta}}=O\left(\frac{1}{\sqrt{N}}\right)\label{eq:Fluctuation Balance}
\end{equation}
requiring $\delta\hat{\mathbf{r}}\approx-\frac{1}{\sqrt{N}}\left(\hat{\mathbf{V}}^{T}\right)^{-1}\hat{\bb{\eta}}$
to leading order.\\

The fluctuation blance equation (Eqn \ref{eq:Fluctuation Balance}),
allows us to derive an analytical expression for the amplitude, structure
and temporal profile of the fluctuations in $\delta\hat{\mathbf{r}}$.
As we show in Appendix \ref{subsec:Appendix-E--Fluctuation Balance}:
\begin{equation}
\hat{\mathbf{C}}\left(\tau\right)=\frac{C\left(\tau\right)}{N}\left(\left[\hat{\mathbf{V}}\hat{\mathbf{V}}^{T}\right]^{-1}-\mathbf{I}\right)\label{eq:Structure of Covariance C_hat}
\end{equation}
where $C\left(\tau\right)\equiv\left\langle \delta r_{i}\left(t\right)\delta r_{i}\left(t+\tau\right)\right\rangle $
is the average single-neuron autocorrelation, which can be calculated
by dynamic mean-field theory given only the balance fixed point, $\hat{\mathbf{r}}^{*}$
(see Appendix \ref{subsec:A2---Dynamic} and SI Fig 1).\\

The total temporal variance in the balance subspace is

\begin{equation}
\left\langle \delta\hat{\mathbf{r}}^{T}\delta\hat{\mathbf{r}}\right\rangle =\frac{C\left(0\right)}{N}\sum_{k=1}^{D}\left(\frac{1}{s_{k}^{2}}-1\right)\label{eq:Total Macroscopic Variance}
\end{equation}

where $s_{k}$ are the singular values of the alignment matrix, $\hat{\mathbf{V}}$
, showing that decreased alignment, as measured by decreased singular
values of the alignment matrix, increases the macroscopic temporal
fluctuations. Furthermore, we can see that as a network approaches
full alignment, $s_{k}\rightarrow1$, the $O\left(\frac{1}{N}\right)$
leading contribution to the net macroscopic fluctuations vanishes,
consistent with our observation (Fig 3A) of $O\left(\frac{1}{N^{2}}\right)$
scaling in fully aligned networks .\\
\\
Our analytical expression (Eqn \ref{eq:Structure of Covariance C_hat})
reveals how misalignment, via $\hat{\mathbf{V}}$, imprints a non-trivial
spatial structure on the fluctuations in the balance subspace. From
Eq. \ref{eq:Structure of Covariance C_hat} (see also Appendix \ref{subsec:Appendix-E--Fluctuation Balance}),
one sees that the eigenvectors of the cross-covariance, $\hat{\mathbf{C}}$,
i.e. the principal components of the balance subspace activity, are
given by the left singular vectors of the alignment matrix, $\hat{\mathbf{V}}$.
The corresponding eigenvalues are determined by the corresponding
singular values of $\hat{\mathbf{V}}$, and given by $\frac{C\left(0\right)}{N}\left(\frac{1}{s_{k}^{2}}-1\right)$.
Thus, the spatial structure of macroscopic fluctuations is entirely
determined by the singular value decomposition of the alignment matrix.
Interestingly, it is independent of $\bb{\Sigma}$, the singular values
of the full structured connectivity, because $\bb{\Sigma}$ multiplies
(Eq. \ref{eq:Fluctuation Dynamics}) both the effective-feedforward
input, $\hat{\bb{\eta}}$, and the balance subspace recurrent input,
$\hat{\mathbf{V}}^{T}\delta\hat{\mathbf{r}}$, and therefore does
not enter the fluctuation balance equation (Eq. \ref{eq:Fluctuation Balance}).\\

Aonther noted feature of Eqn \ref{eq:Structure of Covariance C_hat}
is that $\hat{\mathbf{C}}\left(\tau\right)$ is a product of a $D$-by-$D$
matrix and a scalar temporal profile, thus the time course of macroscopic
fluctuations are identical across the $D$ modes of the balance subspace,
and are given by the average single-neuron autocorrelation function,
$C\left(\tau\right)$.\\

In order to verify our predictions, we construct networks with Gaussian
balance subspace and a heterogeneous set of $\left\{ s_{k}\right\} \in\left[0,1\right]$
(Figure 3D-F, see Appendix \ref{subsec:D2---Heterogeneous Misalignment}
for details). For our choice of parameterization we derive an expression
for the total temporal variance as a function of $\det\hat{\mathbf{V}}=\prod_{k}s_{k}$,
and verify it over an order of magnitude for different instantiations
of $\hat{\mathbf{V}}$ with randomly chosen singular vectors (Fig
3F). For a given choice of $\hat{\mathbf{V}}$, we show that both
our closed-form expressions for the spatial structure (Fig 3D) and
our theoretical prediction of the time-course yield very good predictions
(Fig 3E). \\

For a fixed network size, the macroscopic fluctuations grow with decreased
alignment until at least one singular value, $s_{j}$, is on the order
of magnitude of $\frac{1}{\sqrt{N}}$. As we detail in Appendix \ref{subsec:Appendix-F--NonAlignment},
at that scale, the activity of the corresponding mode is unconstrained
by the leading-order balance equations (Eq. \ref{eq:Linear Bal Eqns}).
At the same time, the recurrent subspace fluctuations (Eqn \ref{eq:Fluctuation Dynamics}),
$\delta\hat{\mathbf{r}}$, are not sufficient to fully cancel the
effective-feedforward fluctuations, $\hat{\bb{\eta}}$, and the fluctuation
balance equation (Eqn \ref{eq:Fluctuation Balance}) cannot be satisfied.
In that situation, the theory derived here breaks down, and macroscopic
synchronous fluctuations with $\delta\hat{\mathbf{h}},\text{\ensuremath{\delta\hat{\mathbf{r}}}}\sim O\left(1\right)$
can emerge. Similar scenarios were studied in \citep{Darshan2020,Hayakawa2020}).\\

The self-consistent solution to the fluctuations describing the fluctuations
in $\hat{\mathbf{h}}$ are beyond the scope of this work.\\

\begin{figure}
\includegraphics{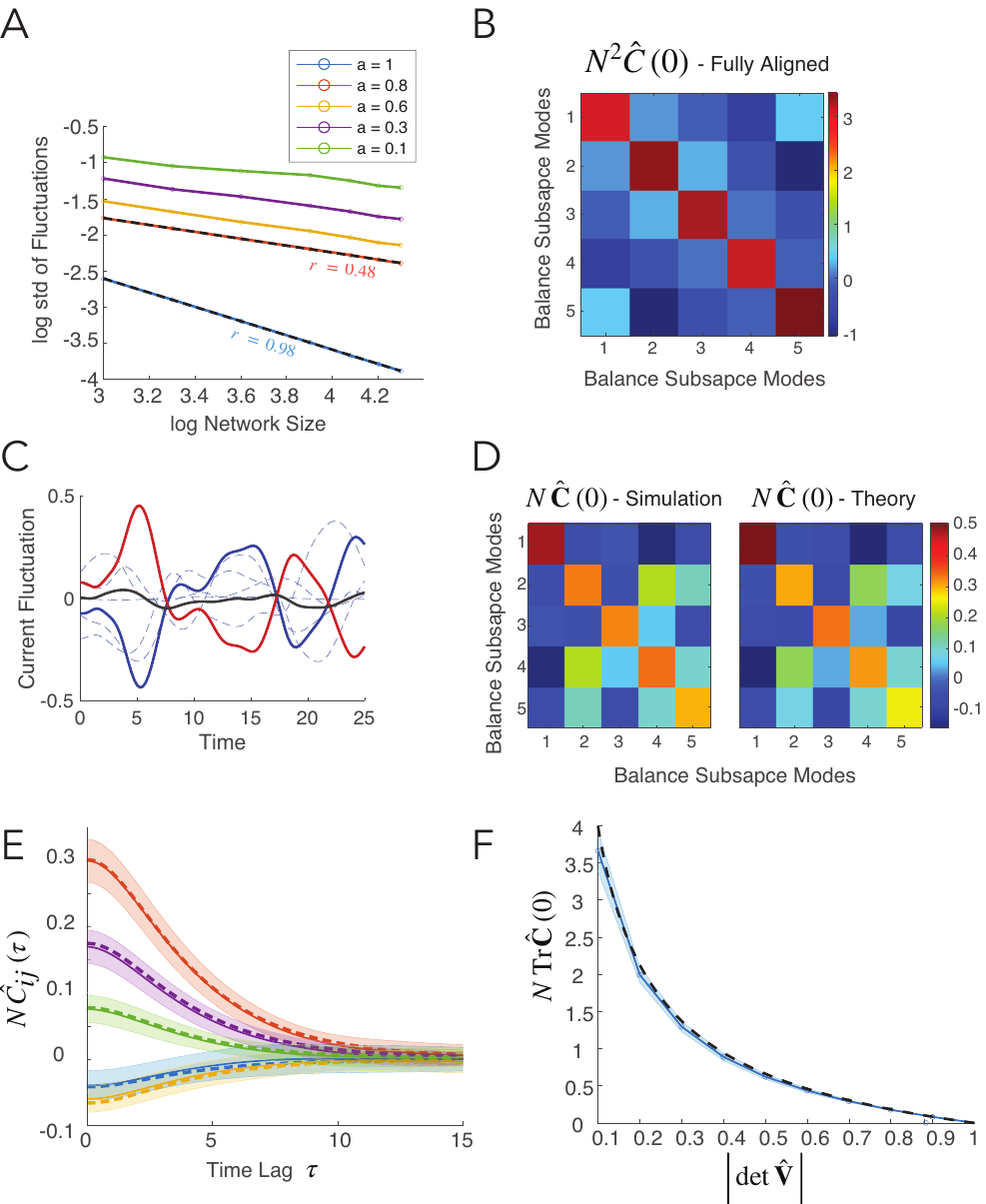}

\caption{\textbf{Partial Misalignment Can Increase Fluctuations in the Balance
Subspace. (A) }Log-log plot of total macroscopic temporal fluctuations
vs network size, i.e. $\frac{1}{2}\log_{10}\protect\trace{}\hat{\mathbf{C}}\left(0\right)$
vs $\log_{10}N$, for varying values of alignment parameter, $a$
($a=1$ is fully aligned. See \ref{subsec:D1---Uniform Misalignment}
for details). Linear fit (dashed line) shows that standard deviation
of fluctuations scales as $\frac{1}{\sqrt{N}}$for networks with incomplete
alignment, as compared to $\frac{1}{N}$ for fully aligned networks.
(\textbf{B) }The\textbf{ }$D$-by-$D$ covariance of firing rate fluctuaitons
in the balance subspace, $\hat{\mathbf{C}}\left(0\right)$, for a
fully aligned network is structureless. Compare (D) for case of incomplete
alignment. \textbf{(C)} Sample trace of fluctuations in input currents
onto a single mode of the balance subspace ($D=5$, $N=8000$ in this
example). $O\left(1\right)$ fluctuations in the ``effective-feedforward''
input from the orthogonal subspace (red) are canceled by fluctuations
in the ``subspace-recurrent input'' from within the balance subspace
(blue, dotted line displays individual input modes and solid line
displays net current fluctuations). Block solid line displays net
flucutations in input current, which are $O\left(\frac{1}{\sqrt{N}}\right)$.\textbf{
(D) }$D$-by-$D$ covariance of firing rate fluctuations in the balance
subspace, displayed here scaled by $N$. (i.e. $N\hat{\mathbf{C}}\left(0\right)$).
Left: single trial simulation. Right: Theory given by Eqn \ref{eq:Structure of Covariance C_hat}.
\textbf{(E) }Time course of cross-correlation function between pairs
of modes, scaled by $N$, $N\hat{C}_{ij}\left(\tau\right)$. Red is
the autocorrelation, other colors display cross-correlation. Dotted
lines show theory (Eqn \ref{eq:Structure of Covariance C_hat}). Solid
lines show simulation results with shaded regions displaying standard
deviation over 20 random realizations of $\mathbf{U}$ for fixed $\hat{\mathbf{V}}$.
In (C) and (D),\textbf{ }$\left|\det\hat{\mathbf{V}}\right|=0.2255$,
the the singular values of $\hat{\mathbf{V}}$ were set to be $\left\{ 0.9,0.825,0.75,0.675,0.6\right\} $
(see Appendix Methods for more details).\textbf{ (F) }Total variance
of firing rate fluctuations in the balance subspace, scaled by $N$,
plotted against the determinant of the alignment matrix, $\hat{\mathbf{V}}$.
Blue shows simulation results with standard deviation over 20 random
realizations. Dotted black line shows theory as given by Eqn \ref{eq:Total Macroscopic Variance}.
In these simulations, the external drive,\textbf{ $\mathbf{f}$},
was rescaled\textbf{ }in order to fix the norm of the balance firing
rates, $\protect\norm{\hat{r}^{*}}$\textbf{ }(see Appendix Methods
for more details) $N=10000$ in all simulations unless otherwise noted.}
\end{figure}

\section{Biologically Relevant Examples\label{sec:Biologically-Relevant-Examples}}

\subsection{Degenerate Excitation-Inhibition Balance\label{subsec:Degenerate-Excitation-Inhibition}}

We first provide a sketch of the application of our generalized balance
framework in the well-known setting of excitation-inhibition balance.
Typically such a network is constructed by randomly and independently
assigning connections to a fraction, $p$, of all pairs of neurons,
with the synaptic weight (and sign) from neuron $j$ to neuron $i$
depending on each of their identities as either excitatory or inhibitory:
\begin{equation}
W_{ij}=\begin{cases}
\pm\frac{J_{\alpha_{i}\alpha_{j}}}{p\sqrt{N}} & \mathrm{with\;\mathrm{prob\;}}p\\
0 & \mathrm{with}\;\mathrm{prob\;}1-p
\end{cases}
\end{equation}
where $\alpha_{i}=I$ for $N_{I}$ inhibitory neurons, and $\alpha_{i}=E$
for the remaining $N_{E}$ excitatory neurons, and the sign of the
weight is corresponding to the pre-synaptic neuron, so that if $\alpha_{j}=E$
($I$) the $j$th column of $W$ is positive (negative). Such a random
binary connectivity matrix can be approximated by a low-rank, deterministic
component with a $2$-by-$2$ block structure, and an additional random
component \citep{Kadmon2015}. The low-rank component, though it is
not symmetric because of the excitation-inhibition structure, is in
general fully aligned: The balance subspace is the two-dimensional
subspace spanned by two block-vectors, one with matching signs and
one with opposing signs: 
\begin{equation}
\mathbf{u}_{0}\sim\left(\begin{array}{c}
+\\
+
\end{array}\right)
\end{equation}
\begin{equation}
\mathbf{u}_{1}\sim\left(\begin{array}{c}
+\\
-
\end{array}\right)
\end{equation}
That is, the balance subspace consists of a ``sum mode'' and a ``difference
mode'' of the excitatory and inhibitory populations. The read-out
performed by the structured connectivity is from the same subspace,
that is, the span of $\left\{ \mathbf{v}_{0},\mathbf{v}_{1}\right\} $
is identical and therefore the alignment matrix is orthonormal.\\

However, consider the situation in which the synaptic weight is independent
of the identity of the post-synaptic neuron: $J_{EI}=J_{II}=J_{I}$
and $J_{EE}=J_{IE}=J_{E}$. Because the average synaptic strengths
in this network do not depend on the identity of the post-synaptic
neuron, the structured component of connectivity is rank one: it is
the outer product of a sum-mode and a difference mode. As Helias et
al \citet{Helias2014} have also shown, this parameterization yields
amplified fluctuations in both the E and I populations. We now show
that this these fluctuations can be characterized as a specific case
of our theory of incomplete alignment.\\
\\
For simplicity, we assume $N_{I}=N_{E}=\frac{N}{2}$, (see Appendix
\ref{subsec:G1---Degenerate E-I with Unequal Pop Size} for general
case). Explicitly we write the structured component of connectivity
as $\frac{\sigma}{\sqrt{N}}\mathbf{u}_{0}\mathbf{v}_{0}^{T}$, according
to our generalized balance formalism, with the following definitions:
\begin{equation}
\mathbf{u}_{0}=\left(\begin{array}{c}
\mathbf{1}\\
\mathbf{1}
\end{array}\right)
\end{equation}
\begin{equation}
\mathbf{v}_{0}=\frac{1}{\sigma}\left(\begin{array}{c}
J_{E}\mathbf{1}\\
-J_{I}\mathbf{1}
\end{array}\right)
\end{equation}
\begin{equation}
\sigma=\sqrt{\frac{J_{E}^{2}+J_{I}^{2}}{2}}
\end{equation}
where $\mathbf{1}$ is the uniform column-vector of length $\frac{N}{2}$.\\

We observe that the alignment matrix is a scalar in this case, and
is given by:
\begin{equation}
\hat{v}=\frac{1}{N}\mathbf{u}^{T}\mathbf{v}=\frac{J_{E}-J_{I}}{2\sigma}
\end{equation}
Inhibition-dominance and network stability will require that $J_{I}\ge J_{E}$.
We assume the external drive is uniform, $\mathbf{f}=\mathbf{1}r_{0}$,
and then the balance equation yields population average firing
\begin{equation}
\hat{r}=-\frac{r_{0}}{\hat{v}\sigma}=\frac{2r_{0}}{J_{I}-J_{E}}
\end{equation}
As the relative strength of inhibition decreases toward parity with
excitation, the alignment between $\mathbf{v}_{0}$ and $\mathbf{u}_{0}$
shrinks: the row-vector, $\mathbf{v}_{0}^{T}$, reads out the difference
between net E and I activity, while the column-vector, $\mathbf{u}_{0}$,
drives E and I equally. To avoid saturation while changing the I-E
ratio, the external drive must shrink to compensate for diminished
alignment, by scaling $r_{0}\propto\hat{v}\sigma=J_{I}-J_{E}$.\\

Applying our theory (Eqn \ref{eq:Structure of Covariance C_hat})
we find that reduced I-E ratio will lead to larger fluctuations of
the population firing rate, $\hat{C}\left(\tau\right)=\left\langle \hat{r}\left(t\right)\hat{r}^{T}\left(t+\tau\right)\right\rangle $:
\begin{equation}
\hat{C}\left(\tau\right)=\frac{C\left(\tau\right)}{N}\left(\frac{J_{E}+J_{I}}{J_{E}-J_{I}}\right)^{2}\label{eq:DegenEI Theory}
\end{equation}

By fixing $J_{E}$ and varying the ratio $\frac{J_{I}}{J_{E}}$ while
scaling $r_{0}$ to maintain the balance firing rate, we confirm this
prediction via simulations in Figure 4F. \\
\\
Note that in the limit of $J_{I}\rightarrow J_{E}$, the expression
for the size of these fluctuations diverges. In that limit, external
drive must be zero in order to avoid saturation, and the size of the
fluctutaions along $\mathbf{u}_{0}$, i.e. coherent fluctuations shared
by the entire population, is $O\left(1\right)$. Our theory breaks
down once the fluctuations in $\hat{r}$ are sufficiently large relative
to $\frac{1}{\sqrt{N}}$, as they begin to significantly impact the
fluctuations in the orthogonal complement \citep{Landau2018,Hayakawa2020}.\\

\subsection{Heterogeneous Out-Degrees}

We now employ our theory of misalignment to study the dynamics of
networks with heterogeneous out-degrees (see Appendix \ref{subsec:G2---Heterogeneous In- and Out-Degrees}
for the case of both heterogeneous out- and in-degrees). Consider
a single inhibitory population in which each neuron $i$ has $K_{i}$
randomly chosen outgoing connections, with each non-zero synapse having
weight $-\frac{J}{\sqrt{N}}$, where $K$ is the average number of
connections per neuron. Just as in the E-I setting, such a random
binary connectivity structure can be approximated by a deterministic
low-rank component and a random component of connectivity. The deterministic
component in this case can be written as:
\begin{equation}
M_{ij}=\frac{J}{\sqrt{N}}\frac{k_{j}}{N}
\end{equation}
where we have defined the relative out-degrees, $k_{i}\equiv\frac{K_{i}}{K}$.\\

This deterministic $\mathbf{M}$ is a rank-one matrix given by $\mathbf{M}=\sigma\mathbf{u}_{0}\mathbf{v}_{0}^{T}$,
with the following definitions: 
\begin{equation}
\mathbf{u}=\mathbf{1}
\end{equation}
\begin{equation}
\mathbf{v}=\frac{\mathbf{k}}{\sqrt{\left\langle k^{2}\right\rangle }}
\end{equation}
\begin{equation}
\sigma=\sqrt{\left\langle k^{2}\right\rangle }Jp
\end{equation}
where $p\equiv\frac{K}{N}$, $\left\langle k^{2}\right\rangle $ is
the mean-square of the relative out-degrees, and $\mathbf{1}$ is
the column vector of all ones.\\

The alignment matrix is scalar in this case as well and is given by
\begin{equation}
\hat{v}=\frac{\mathbf{1}^{T}\mathbf{k}}{N}\frac{1}{\sqrt{\left\langle k^{2}\right\rangle }}=\frac{1}{\sqrt{\left\langle k^{2}\right\rangle }}
\end{equation}
Thus we see that the extent of alignment, and therefore the strength
of ``subspace reccurence'' decreases with increasing breadth of
the out-degree distributions\\
\\
The balance equation gives
\begin{equation}
\hat{r}=-\frac{r_{0}}{\hat{v}\sigma}=\frac{r_{0}}{Jp}
\end{equation}
such that the balance firing rates in the population are unaffected
by broadening of the out-degree. This is intuitive because the mean
recurrent input is not expected to depend on the breadth of the out-degree.\\
\\
Nevertheless, broad a out-degree distribution increases the extent
of coherent fluctuations, as decreased alignment means an increasing
of ``effective-feedforward'' connectivity from the orthogonal complement
to the balance subspace. Concretely, we have:
\begin{equation}
\hat{C}\left(\tau\right)=\frac{C\left(\tau\right)}{N}\left(\left\langle k^{2}\right\rangle -1\right)\label{eq:Outdegree Theory}
\end{equation}

Note that $\left\langle k^{2}\right\rangle -1$ is exactly the variance
of the relative out-degrees. The increase of correlations with broader
out-degree was observed in \citet{Roxin2011}.\\

We find that broader out-degree distribution leads to larger coherent
fluctuations and confirm this prediction via simulations (Fig 4F).
It is worth noting that the structure of $\mathbf{V}_{\perp}$ does
not enter the mean-field theory, beyond the assumption that $\bb{\eta}=\frac{1}{\sqrt{N}}\mathbf{V}_{\perp}^{T}\delta\mathbf{r}$
is an independent mean-zero Gaussian process. In practice, we generate
out-degrees from a log-normal distribution and find that the simulations
fit the theory well within a broad range of out-degree variability,
although the fit worsens as the variability increases, as presumably
the Gaussian assumption is violated.

\begin{figure}
\includegraphics{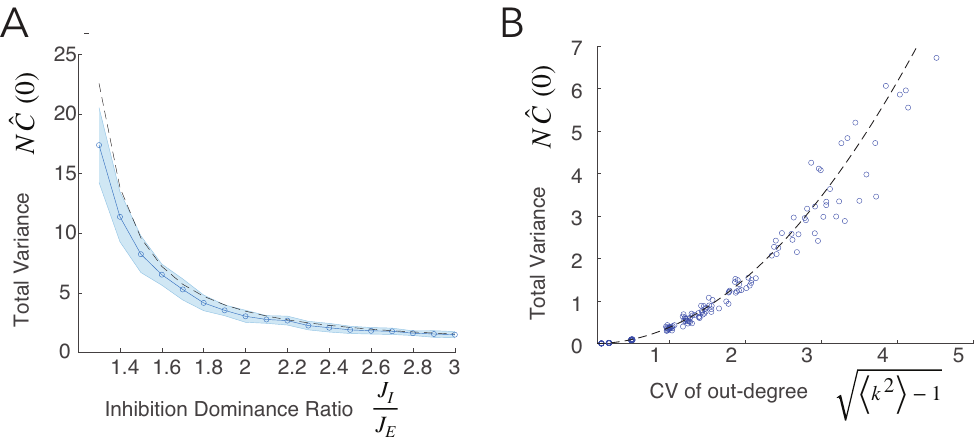}

\caption{\textbf{Biologically Relevant Examples of Incomplete Alignment Yiedling
Low-Dimensional Fluctuations (A) }E-I network with degenerate weight
parameters in the sense that the strength of both E and I synapses
are independent of the identity of the post-synaptic neuron. Variance
of population firing rate over time, scaled by $N=20000$ in this
simulation. Dotted black line shows theory (Eqn \ref{eq:DegenEI Theory}).
Shaded region shows standard error of the mean over 20 trials. \textbf{(B)}
Inhibitory network with heterogeneous out-degree. Variance of the
population firing rate over time, scaled by $N$, as a function of
the coefficient of variation (CV) of the the out-degree. Dotted line
shows theory (Eqn \ref{eq:Outdegree Theory}). Circles show single
trial simulations. Out-degrees were drawn from a log-normal distribution.
In-degrees were taken to be uniform. $N=10000$.}
\end{figure}

\section{Discussion}

We have generalized the theory of dynamic excitation-inhibition balance
to networks with arbitrary low-rank structured connectivity, $\mathbf{M}=\frac{1}{\sqrt{N}}\mathbf{U}\bb{\Sigma}\mathbf{V}^{T}$,
together with an additive random component of connectivity. We decompose
the network dynamics into the ``balance subspace'', given by $\span{\mathbf{U}}$,
which receives strong structured recurrent input, and its orthogonal
complement within which the recurrent input is driven by random connectivity.
The core insight of our theory is that the macroscopic, low-dimensional
dynamics in the balance subspace are determined by the alignment matrix,
$\hat{\mathbf{V}}=\frac{1}{N}\mathbf{U}^{T}\mathbf{V}$, between left-
and right-singular vectors. We derive linear balance equations for
the activity in the balance subspace which are a generalization of
the two balance equation in excitation-inhibition networks. These
balance equations are independent of the underlying statistics of
the balance subspace itself, $\mathbf{U}$, but rather depend only
on the alignment, $\hat{\mathbf{V}}$, and the singular values, $\bb{\Sigma}$,
which scale the recurrent connectivity in each of the balance subspace
modes.\\

We observe that $\hat{\mathbf{V}}$quantifies the alignment between
row-space and column-space of the structured connectivity, that is,
between the the read-out and feed-back subspaces. If the read-out
subspace is identical to the feed-back subspace, then $\hat{\mathbf{V}}$
is orthonormal. If $\hat{\mathbf{V}}$ is not an orthonormal matrix
then the structured connectivity is not fully aligned. In this case
the structured connectivity can be decomposed to a subspace-recurrent
component that reads-out network activity strictly from within the
balance subspace, and an effective-feedforward component that reads-out
network activity from the orthogonal complement and projects it to
the balance subspace. \\

We show that increased misalignment increases coherent, low-dimensional
fluctuations in the balance subspace. This is due to the strengthening
of the effective-feedforward component of the structured connectivity,
which reads-out microscopic fluctuations from the orthogonal complement
and projects them to the balance subspace. It is important to emphasize
that the fluctuating input currents projected from the orthogonal
complement to the balance subspace are of $O\left(1\right)$, but
that balance leads to a dynamic canceling of these fluctuating currents
to leading order. This cancelation is manifested by a set of fluctuation-balance
equations that enable us to determine the size, structure, and temporal
profile of firing rate fluctuations in the balance subspace, $\hat{\mathbf{C}}\left(\tau\right)$,
as a function of $\hat{\mathbf{V}}$. Despite the suppression of fluctuations
via dynamic balance, the magnitude of coherent fluctuations due to
misalignment is an order of magnitude larger than under fully aligned
connectivity. These fluctuations increase with misalignment as a function
of the singular values of $\hat{\mathbf{V}}$, as the subspace-recurrent
feedback is weakened requiring larger macroscopic fluctuations in
order to cancel the input fluctuations from the orthogonal complement.
Because they are driven by the microscopic fluctuations in the orthogonal
complement, the temporal profile of these shared macroscopic fluctuations
is identical to the average temporal profile of single-neuron fluctuations.
The spatial structure of the fluctuations within the $D$-dimensional
balance subspace is given by the left-singular vectors of the alignment
matrix. \\

Note that in the limit where at least one of the singular values of
$\hat{\mathbf{V}}$ is zero, fluctuation balance cannot be achieved
along the corresponding modes. Any external drive into these modes
will drive saturation, but if there is no external drive to these
modes, then the fluctuations in the orthogonal complement will be
propagated to this mode via the structured connectivity in a fully
feed-forward manner, without being suppressed by dynamic balance.
This can lead to $O\left(1\right)$ shared fluctuations as studied
in \citep{Landau2018}.\\

We have studied continuous rate-neuron models, for their analytical
tractability. However, we expect that our theory will hold qualitatively
for spiking-neuron network models as well. This is because we find
that despite the elevated shared fluctuations, networks with incomplete
alignment exhibit asynchronous dynamics as long as the singular values
of $\hat{\mathbf{V}}$are non-zero. Thus we expect that spiking networks
with strong low-rank connectivity structure will operate in an asynchronous,
irregular regime that is well-described by a mean-field theory \citep{Brunel2000}
which can be etended to include the impact of misaligned structure
as we have done here.\\

We have studied the structure of temporal fluctuations but our approach
can be readily extended to study the structure of quenched variability
over multiple realizations of the random connectivity. We expect that
incomplete alignment will amplify quenched fluctuations in a similar
manner.\\

Note that throughout this work we assume that the external drive vector,
$\mathbf{\hat{\mathbf{f}}}$, is such that enables a balanced fixed
point. This requires that the external drive be almost entirely contained
in the balance subspace (up to $O\left(\frac{1}{\sqrt{N}}\right)$
projections). The impact of misaligned external drive is to effectively
weaken the input to the balance subspace, thus shifting the balance
fixed point and eventually suppressing chaos and yielding saturation,
as we showed in \citep{Landau2020}. In addition, a balanced fixed
points imposes further constrains on $\hat{\mathbf{f}}$. For example,
in standard E-I balance networks, external drive onto the E population
must be sufficiently stronger that that onto the I population in order
to maintain non-negative firing rates \citep{VanVreeswijk1998,Baker2020}.
Describing these constraints in the general low-rank setting is beyond
the scope of this work. \\

We note that the random component of connectivity may have statistical
structure, for example different cell-types may have different overall
variability of their synaptic strengths yielding a block-structure
on the variances \citep{Aljadeff2015,Kadmon2015,Schuessler2020}.
Additionally, the low-rank structure may be correlated with the random
component of connectivity \citep{Schuessler2020}. These types of
correlations in the connectivity may have non-trivial impact in the
regime of dynamic balance which may be of interest for future study.\\

\subsection*{Relationship to Previous Models of Dynamic Balance}

Our framework of generalized dynamic balance unifies much previous
work. For example the typical E-I balance network has rank-$2$ deterministic
structure, in which both the column-space and row-space consist of
a ``sum'' mode and a ``difference'' mode, and therefore the structure
is fully aligned. Some E-I network models (e.g. \citep{Brunel2000,Ostojic2014}),
however, use a degenerate block-structure by constraining the average
weights to be independent of post-synaptic population. As we have
shown here, in this case, the structured connectivity has rank $1$
and is only partially aligned. The balance subspace is a sum mode,
$\mathbf{u}_{0}\propto\mathbf{1}$, while the read-out mode, $\mathbf{v}_{0}$,
is a difference mode. As the relative strength of inhibition decreases,
the recurrent connectivity becomes less alignedand yields larger and
larger coherent fluctuations. These increased fluctuations in the
case of degenerate E-I structure have been previously studied in \citep{Helias2014}.
As suggested in the discussion in \citep{Helias2014}, even stronger
correlations arrise in the case of a ``doubly-degenerate'' E-I structure,
in which in addition to being independent of post-synaptic population,
the average weights are set to zero. Such a case was studied in \citep{Hayakawa2020}
and corresponds to the limit where $\hat{\mathbf{V}}=0$ in our formalism.\\

E-I networks with distance dependent connectivity have been the focus
of a number of past studies \citep{Vreeswijk2005,Rosenbaum2014,Darshan2020,Ebsch2020}.
In the typically studied setting, each of the E and I connectivity
profiles share the same periodic boundary conditions and are constrained
to finite spatial frequency modes, and therefore both the column-space
and row-space are spanned by a concatenated Fourier basis of E and
I cells, and therefore$\hat{\mathbf{V}}$ is orthonormal. \citep{Darshan2020}
present cases in which E-I networks with distance-dependent connectivity
have a singular $\hat{\mathbf{V}}$. For example, they consider the
case in which the I-to-E connectivity has spatial dependence while
all other connectivity profiles are spatially uniform. In our formalism,
the column-space (columns of $\mathbf{U}$) no longer includes Fourier
modes of the I population, but rather consists of Fourier modes of
the E population concatenated with the zero vector over the I population.
On the other hand the row-space (columns of $\mathbf{V})$ does not
include Fourier modes from the E population, but rather, consists
of I-population Fourier modes concatenated with the zero vector over
the E population, and these modes are entirely orthogonal to $\mathbf{U}$
so that they are fully feed-forward. Thus spatially correlated fluctuations
are propagated from the I population to the E population, where they
are not canceled. As further discussed in \citep{Darshan2020}, the
situation is different if the I population has distant-dependent connectivity
internally as well. In our formalism this would mean that while columns
of $\mathbf{V}$ would still not include Fourier modes of E, the columns
of $\mathbf{U}$ would once again consist of the concatenated E and
I Fourier modes. Thus, this is an example of partial alignment. Therefore,
the spatially correlated fluctuations in the presence of I-to-I distance
dependent are an order of magnitude smaller than the purely feed-forward
case, they are still an order of magnitude larger than in the fully
aligned case in which the E population also has distance-dependent
projections (whether to itself or to the I population).\\

Networks with heterogeneous in-degrees have been previously shown
to exhibit broken balance \citep{Pyle2016,Landau2016a}. That result
can be undestood in a straightforward manner in our framework of generalized
balance: if the in-degrees from the external drive are not a linear
combination of E and I in-degrees, then the external drive is not
fully aligned with the balance subspace and there will be strong external
drive to the orthogonal complement which will drive saturation \citep{Landau2020}.
In addition to heterogeneous in-degrees, here we have studied the
impact of heterogeneous out-degrees on dynamic balance. We show that
broad out-degree distributions in balanced networks are a form of
incomplete alignment and result in increased coherent fluctuations.
A similar phenomenon was observed numerically in \citep{Roxin2011}.
We provide an analytical expression for the size of fluctuations as
a function of the breadth of the out-degrees, and verify it numerically.
We furthermore show that negative correlations between in- and out-degrees
will further amplify the shared fluctuations (Appendix\ref{subsec:G2---Heterogeneous In- and Out-Degrees})\\

Other work has shown that balance networks can exhibit shared variability
if they have shared fluctuations in their external input \citep{Darshan2017,Rosenbaum2016}.
Our work assumes a constant external input.

\subsection*{Conclusion}

Previous studies have explored particular examples of low-rank deterministic
structure in balanced networks, most often via distance-dependent
connectivity \citep{Vreeswijk2005,Rosenbaum2014,Rosenbaum2016,Darshan2020}
or sub-population structure \citep{Kadmon2015,Darshan2017}. In such
low-rank structures, the column-space (the span of the left-singular
vectors) is typically identical to the row-space (the span of the
right-singular vectors). We call such networks ``fully aligned'',
and study the more general situation of partial alignment in which
the row-space is not entirely contained in the column-space of the
low-rank matrix. We show that such incomplete alignment can have qualitative
impact on network dynamics. The key feature of the structured connectivity
in our analysis is the alignment matrix, comprised of the overlaps
between left- and right-singular vectors.\\

Low-rank structured connectivity may reflect different cell-types,
distance-dependence, functional connectivity, as well as heterogeneity
between neurons \citep{Vreeswijk2005,Rosenbaum2014,Landau2016a,Darshan2020}.
Such a generalization may be important for incorporating biological
realism into balance-network models. From another perspective, low-rank
structure has been of recent interest in designing networks that perform
specific computations \citep{Sussillo2009,Mastrogiuseppe2017a}. Our
work studies a new regime where low-rank connectivity is strong, and
suggests a bridge between networks designed for computation and biological
networks exhibiting dynamic balance.\\

We have developed a generalized theory of dynamic balance, which both
unifies previous studes and reveals new results. Our generalization
expands the study of dynamic balance to a broad class of low-rank
structures -- those with only partial alignment between column-space
and row-space. These structures have previously only appeared in particular
cases, but in our framework they appear as the general case of low-rank
connectivity. We show that incomplete alignment generates coherent
fluctuations via effective-feed-forward propagation from a high-dimensional
subspace with microscopic chaos to a low-dimensional, balance subspace.
We derive a set of fluctuation balance equations that provides an
analytical solution for the structure of coherent fluctuations in
the balance subspace.\\

This theory may find relevance well beyond neuroscience. Recent studies
of complex systems attempt to explore the relationship between structure
and dynamics in a variety of real-world networks \citep{Barzel2013,Hens2019}.
Many of these studies limit the strength of interaction between units
($\sim\frac{1}{N}$) in order to facilitate mean-field approaches.
The theory of excitation-inhibition balance studies a regime of strong
interactions ($\sim\frac{1}{\sqrt{N}}$) but until now its application
has remained limited to neuroscience because of the excitation-inhibition
structure (Dale's Law). Our generalized framework of dynamic balance
may be relevant for any setting with strongly interacting units, whether
biological, social, or technological networks.\\

\part*{Acknowledgments}

We thank David Hansel and Yoram Burak for useful comments on previous
versions of this manuscript. H.S. was funded by the Swarz Program
in Theoretical Neuroscience at Harvard, the Gatsby Charitable Foundation,
and NIH grant NINDS (1U19NS104653).

\part*{\protect\pagebreak Appendix}

\subsection*{Appendix A - Self-Consistency of the Balance Solution\label{subsec:Appendix-A--Self-consistency}}

As discussed in the bain text, the macroscopic dynamics in the balanced
subspace (Eqn \ref{eq:Balance Subspace Dynamics}) admit a balanced
fixed point governed by the $D$ linear balance equations: $\bb{\Sigma}\hat{\mathbf{V}}^{T}\hat{\mathbf{r}}^{*}+\hat{\mathbf{f}}\approx0$
(Eqn \ref{eq:Linear Bal Eqns} in the main text). Balance is achieved
when the firing rates in the balance subspace satisfy this equation
up to a finite-size correction of $\frac{1}{\sqrt{N}}$. These firing
rates, $\hat{\mathbf{r}}$, are given by the components of the full
population firing rate vector, $\mathbf{r}=\phi\left(\mathbf{h}\right)$,
along $\mathbf{U}$:
\begin{equation}
\hat{\mathbf{r}}^{*}=\frac{1}{N}\mathbf{U}^{T}\phi\left(\mathbf{U}\hat{\mathbf{h}}^{*}+\mathbf{h}_{\perp}\right)
\end{equation}
As we shall see, these $D$ equations constrain both the residual
fields in the balance subspace, $\hat{\mathbf{h}}^{*}$ , as well
as the statistics of the microscopic, local dynamics in the orthogonal
complement, $\mathbf{h}_{\perp}$. \\

The dynamical state in the orthogonal complement can be either a stable
fixed point (FP) or chaos. Given a fixed $\hat{\mathbf{h}}^{*}$,
those dynamics are described by a mean-field theory that predicts
that the microscopic degrees of freedom, $h_{\perp_{i}}$, are be
described as independent, identical, mean-zero Gaussians, either fixed
in time or fluctuating with a monotonically decaying autocorrelation.
The mean-field theory follows previous work (e.g. \citep{Sompolinsky1988,Kadmon2015}),
and we detail it for our setting below in Appendix \ref{subsec:Appendix-B--MF in the Orthogonal}.
The result of the mean-field theory is that if balance is achieved,
avoiding saturation, then there is a FP with Gaussian statistics that
transitions to chaos for sufficiently strong random component of connectivity.
The expression for the variance in the orthogonal subspace, $\Delta_{0}=\left\langle h_{\perp_{i}}^{2}\right\rangle $,
depends only on $\hat{\mathbf{h}}^{*}$ and $g$. In the FP regime,
$\Delta_{0}$ is the spatial disorder and is given by a single implicit
equation, while in the chaotic regime, $\Delta_{0}$ averages both
spatial and temporal disorder and is constrained by a pair of equations
together with the spatial variance of the single-neuron long-time
averages. \\
\\
Given, $\Delta_{0}$, $\hat{\mathbf{r}}$ is given by first averaging
over the Gaussian component in the orthogonal subspace and then projecting
the result onto balance subspace: 
\begin{equation}
\hat{\mathbf{r}}^{*}=\frac{1}{N}\mathbf{U}^{T}\left\langle \phi\left(\mathbf{U}\hat{\mathbf{h}}^{*}+\sqrt{\Delta_{0}}z\right)\right\rangle _{Dz}\label{eq:MF Subspace Firing Rates}
\end{equation}
where $\left\langle \right\rangle _{Dz}$ is an integral over the
standard normal measure, $Dz=\frac{dz}{2\pi}\ex{-\frac{z^{2}}{2}}$.
Combining the balance balance requirement $\hat{\mathbf{r}}^{*}\approx\hat{\mathbf{r}}^{bal}$
(Eqn \ref{eq:Linear Bal Eqns}) gives us a set of $D$ implicit equations,
which given $\Delta_{0}$, determine $\hat{\mathbf{h}}^{*}$: 
\begin{equation}
\frac{1}{N}\mathbf{U}^{T}\left\langle \phi\left(\mathbf{U}\hat{\mathbf{h}}^{*}+\sqrt{\Delta_{0}}z\right)\right\rangle _{Dz}=-\hat{\mathbf{V}}\bb{\Sigma}^{-1}\hat{\mathbf{f}}\label{eq:Full Balance Eqns, Non-Linear}
\end{equation}

We stress that this Gaussian mean-field equation for self-consistency
in the balance subspace holds regardless of whether the dynamics in
the orthogonal subspace are in the FP or chaotic regime, but the total
variance in the orthogonal subspace, $\Delta_{0}$, must also be found
in a manner self-consistent with $\hat{\mathbf{h}}^{*},$as we show
below.\\

\subsection*{Appendix B - Mean-Field Theory in the Orthogonal Subspace\label{subsec:Appendix-B--MF in the Orthogonal}}

We now detail the mean-field theory describing the dynamics in the
orthogonal subspace for both fixed point and chaos. We will assume
an approximate fixed point in the balance subspace, $\hat{\mathbf{h}}^{*}$,
that enables the corresponding firing rates, $\hat{\mathbf{r}}^{*}$,
to satisfy the $D$ balance equations (Eqn \ref{eq:Linear Bal Eqns}).\\

We further assume full external alignment, as we have throughout the
main text, so that the external drive, $\mathbf{f}$, does not project
into the orthogonal subspace. The dynamics are given by:
\begin{equation}
\drdt{\mathbf{h}_{\perp}}=-\mathbf{h}_{\perp}+\frac{g}{\sqrt{N}}\mathbf{J}_{\perp}\mathbf{r}
\end{equation}
where, we remind the reader, $\mathbf{X}_{\perp}=\left(\mathbf{I}-\frac{1}{N}\mathbf{U}\mathbf{U}^{T}\right)\mathbf{X}$
is the orthogonal compliment of the vector or matrix of column-vectors,
$\mathbf{X}$, and the vector of firing rates is given by $\mathbf{r}=\phi\left(\mathbf{h}\right)$.
Through the non-linearity, $\phi$, the dynamics in the orthogonal
complement depend on the dynamics in the balance subspace, $\hat{\mathbf{h}}^{*}$.

\subsubsection*{\label{subsec:A.1.-Fixed-Point}B1 Fixed Point and its Stability}

The fixed point equation in the orthogonal subspace is 
\begin{equation}
\mathbf{h}_{\perp}^{*}=\mathbf{J}_{\perp}\mathbf{r}^{*}
\end{equation}

Given a fixed $\hat{\mathbf{h}}^{*}$, we follow the mean-field theory
presented, for example in \citep{Kadmon2015}, which treats the recurrent
drive due to the random connectivity, as independent, identical, mean-zero
Gaussians. This theory assumes that due to the random connectivity,
$\left(\sum_{j}^{N}J_{\perp_{ij}}r_{j}\right)^{2}\approx\sum_{j}J_{\perp_{ij}}^{2}r_{j}^{2}\approx N\left\langle J_{ij}^{2}\right\rangle \left\langle r_{j}^{2}\right\rangle $,
which therefore determines the spatial variance at the fixed point,
$\Delta_{0}\equiv\left\langle h_{\perp_{i}}^{2}\right\rangle $, to
be
\begin{equation}
\Delta_{0}=g^{2}\left\langle r_{j}^{2}\right\rangle 
\end{equation}
The mean-square firing rates, $\left\langle r_{j}^{2}\right\rangle =\left\langle \phi^{2}\left(h_{j}\right)\right\rangle $,
must be found by averaging over the network, where $h_{j}$ has a
component in the balance subspace and an additional Gaussian component
with variance $\Delta_{0}$. Because the two components are independent,
we can perform the average over the Gaussian randomness before averaging
over the population, and thus we arrive at an implicit mean-field
fixed-point equation for $\Delta_{0}$: 
\begin{equation}
\Delta_{0}=g^{2}\frac{\mathbf{1}^{T}}{N}\left\langle \phi^{2}\left(\mathbf{U}\hat{\mathbf{h}}^{*}+\sqrt{\Delta_{0}}z\right)\right\rangle _{Dz}\label{eq:Delta Implicit Equation-1}
\end{equation}

where $\left\langle \right\rangle _{Dz}$ means averaging over the
standard normal measure $Dz=\mathrm{e}^{-\frac{z^{2}}{2}}dz$.\\

Given a fixed $\hat{\mathbf{h}}^{*}$, the Jacobian matrix for the
stability of the fixed point in the orthogonal complement is given
by $J_{\perp_{ij}}\phi'\left(h_{j}\right)$. As we have done previously,
we assume that $h_{j}$ is independent of $J_{\perp_{ij}}$, and this
Jacobian matrix is a random matrix with column-wise variance. The
support of the eigenvalues of such a matrix is identical that of a
random matrix with uniform variance that is given by the average of
the column-wise variances \citep{Ahmadian2015}. Thus, the effective
gain at the fixed point, which is given by the maximal real part of
the eigenvalues of the Jacobian, is given by $g_{eff}^{2}=\left\langle J_{\perp_{ij}}^{2}\right\rangle \left\langle \phi'\left(h_{j}\right)^{2}\right\rangle $,
which can be calculated as:
\begin{equation}
g_{eff}^{2}=g^{2}\frac{\mathbf{1}^{T}}{N}\left\langle \phi'^{2}\left(\mathbf{U}\hat{\mathbf{h}}^{*}+\sqrt{\Delta_{0}}z\right)\right\rangle _{Dz}\label{eq:Effective Gain-1}
\end{equation}
The fixed point in the orthogonal complement, $\mathbf{h}_{\perp}^{*}$,
will be stable for $g_{eff}<1$, and for $g_{eff}>1$ the microscopic
dynamics in the orthogonal compliment will be chaotic. Those dynamcis
are described by a dynamic mean-field theory (DMFT), detailed in the
next section.\\

The stability calculation here is equivalent to considering perturbations
within the orthogonal complement, with the balance subspace held fixe.
A complete treatment of stability should consider arbitrary perturbations,
in both the orthogonal complement and the balance subspace, following
\citep{Kadmon2015,Mastrogiuseppe2017a}.

\subsubsection*{\label{subsec:A2---Dynamic}B2 - Dynamic Mean-Field Theory of Chaos
in the Orthogonal Subspace}

In the chaotic regime, the input to the orthogonal subspace can still
be considered Gaussian but its temporal statistics must be derived,
that is we seek to find the autocorrelation:
\begin{equation}
\Delta\left(\tau\right)\equiv\left\langle h_{i}^{\perp}\left(t\right)h_{i}^{\perp}\left(t+\tau\right)\right\rangle 
\end{equation}
The dynamics in the orthogonal subspace can be represented by a single
stochastic differential equaiton:

\begin{equation}
\drdt{h_{i}^{\perp}}=-h_{i}^{\perp}+\eta_{i}
\end{equation}
where $\eta_{i}\left(t\right)=\sum_{j}J_{ij}^{\perp}r_{j}\left(t\right)$.
Again, due to the randomness of the connectivity, $\eta_{i}$ has
mean-zero over neurons and time, and the average autocorrelation of
$\eta_{i}$ is a scaled version of the average autocorrelation of
$r_{i}$:
\begin{equation}
\left\langle \eta_{i}\left(t\right)\eta_{i}\left(t+\tau\right)\right\rangle =g^{2}C_{T}\left(\tau\right)
\end{equation}
where we have written 
\begin{equation}
C_{T}\left(\tau\right)\equiv\left\langle \phi\left(h_{i}\left(t\right)\right)\phi\left(h_{i}\left(t+\tau\right)\right)\right\rangle 
\end{equation}
 for the average autocorrelation of the firing rates. Note that the
notation $\left\langle \;\right\rangle $ now denotes averaging over
both time and neurons. Additionally, note that we have inserted the
notation $C_{T}$ for the total autocorrelation in order to differentiate
from the mean single-neuron temporal autocorrelation ($C\left(\tau\right)=C_{T}\left(\tau\right)-\left\langle \phi\left(h_{i}\right)\right\rangle ^{2}$)
introduced in the main text in Section \ref{sec:Incomplete-Align Fluctuations},
Equation \ref{eq:Structure of Covariance C_hat}.\\

To compute $C_{T}\left(\tau\right)$ we first write $h_{i}=\mathbf{u}_{i}^{T}\hat{\mathbf{h}}^{*}+h_{i}^{\perp}$,
where $\mathbf{u}_{i}$ is the $i$th row of the left-singular vector
matrix, and gives the projection of neuron $i$ in the balance subspace.
Next we rewrite the two correlated Gaussians, $h_{i}^{\perp}\left(t\right)$
and $h_{i}^{\perp}\left(t+\tau\right)$, via three independent Gaussians,
one of which contributes the correlated component:
\begin{equation}
h_{i}^{\text{\ensuremath{\perp}}}\left(t\right)=\sqrt{\Delta_{0}-\Delta\left(\tau\right)}x_{1}+\sqrt{\Delta\left(\tau\right)}y
\end{equation}
\begin{equation}
h_{i}^{\perp}\left(t+\tau\right)=\sqrt{\Delta_{0}-\Delta\left(\tau\right)}x_{2}+\sqrt{\Delta\left(\tau\right)}y
\end{equation}
where we have introduced $\Delta_{0}\equiv\Delta\left(0\right)$ as
the total variance. These three Gaussians need to be integrated over,
and then the balance subspace structure averaged to yield
\begin{equation}
C_{T}\left(\tau\right)=\frac{1}{N}\sum_{i}\left\langle \left\langle \phi\left(\mathbf{u}_{i}^{T}\hat{\mathbf{h}}^{*}+\sqrt{\Delta_{0}-\Delta\left(\tau\right)}x+\sqrt{\Delta\left(\tau\right)}y\right)\right\rangle _{Dx}^{2}\right\rangle _{Dy}\label{eq:C(tau) general case}
\end{equation}

Thus given the autocorrelation, $\Delta\left(\tau\right)$, in the
orthogonal complement, we have an expression for the average single
neuron firing rate autocorrelation, $C_{T}\left(\tau\right)$. Next,
following the standard DMFT approach (\citep{Sompolinsky1988,Kadmon2015})
we write an implicit differential equation that determines the autocorrelation
self-consistently:
\begin{equation}
\left(1-\pddr{}{\tau}\right)\Delta\left(\tau\right)=g^{2}C_{T}\left(\tau\right)\label{eq:DMFT diff-eqn}
\end{equation}
The total variance, $\Delta_{0}$, is the initial condition of Eqn
\ref{eq:DMFT diff-eqn} and must be found self-consistently along
with the conditions $\pdr{}{\tau}\Delta\left(0\right)=0$ and $\pddr{}{\tau}\Delta\left(\infty\right)=0$.
As detailed in \citep{Sompolinsky1988,Kadmon2015}, the differential
equation (Eqn \ref{eq:DMFT diff-eqn}) can be re-expressed as a one-dimensional
dynamics under a potential energy. Given $\hat{\mathbf{h}}$, the
initial condition, $\Delta_{0}$, can be found by using the requirement
that the potential energy at the initial condition equals its value
at $\tau\rightarrow\infty$, together with the fact that $\Delta\left(\infty\right)=g^{2}C_{T}\left(\infty\right)$.
Therefore, in practice, given the fixed point in the balance subspace,
$\hat{\mathbf{h}}^{*}$, the total variance in the orthogonal subspace,
$\Delta_{0}=\left\langle \left(h_{i}^{\perp}\right)^{2}\right\rangle $,
and the spatial variance of the time-averages, $\Delta\left(\infty\right)=\frac{1}{N}\sum_{i}\left\langle h_{i}^{\perp}\right\rangle _{t}^{2}$,
are found via a pair of coupled equations. The fixed point in the
balance subspace depends on $\Delta_{0}$ in turn, via the balance
equations (Eqn \ref{eq:Full Balance Eqns, Non-Linear}). \\

In sum, the mean-field characterization of the system determines $\hat{\mathbf{h}}^{*}$
and $\Delta_{0}$ via either $D+1$ equations (in the FP regime) or
$D+2$ equations (in the chaotic regime). When these equations are
satisfied the network generalizes the dynamic balance of excitation-inhibition.
This balance is dynamic in the sense that without fine-tuning, the
macroscopic firing rates in the balance subspace self-adjust to cancel
the strong external drive. Depending on the single-neuron transfer
function and the strength of the random component of connectivity,
balance can take the form of either a stable fixed-point, or the more
familiar balanced state of chaotic dynamics with local fluctuations
propagating in the orthogonal complement.\\

\subsection*{Appendix C - The Case of Gaussian Structured Connectivity}

Here we study the mean-field theory for the case in which the structured
connectivity is Gaussian. In particular, the elements of $\mathbf{U}$
are taken to be i.i.d. by $\xnorm\left(0,1\right)$. The Gaussianity
construction greatly simplifies the mean-field expression for $\hat{r}_{k}$
(Eqn \ref{eq:MF Subspace Firing Rates}). In the general case, calculating
$\hat{r}_{k}$ requires first averaging $r_{i}=\phi\left(\mathbf{u}_{i}^{T}\hat{\mathbf{h}}+h_{i}^{\perp}\right)$
over the variability in the orthogonal subspace by a Gaussian integral
for each neuron $i$ (where $\mathbf{u}_{i}$ is the $i$th row of
$\mathbf{U}$ as above), and then computing a weighted average over
the structure of the $k$th column of $\mathbf{U}$. That is
\begin{equation}
\hat{r}_{k}=\frac{1}{N}\sum_{i=1}^{N}u_{ik}\left\langle \phi\left(\sum_{l=1}^{D}u_{il}\hat{h}_{l}+\sqrt{\Delta_{0}}z\right)\right\rangle _{Dz}
\end{equation}
In the Gaussian case, the other modes $l\neq k$ contribute Gaussian
variablilty which can simply be added to the variability from the
orthogonal subspace, and then $u_{ik}$ can be averaged over as an
additional Gaussian. Therefore $\hat{r}_{k}$ reduces to a double
Gaussian integral: one integral for the weighted average over the
structure of the $k$th mode, which is coupled to $\hat{h}_{k}$,
and a second Gaussian that combines the remaining $D-1$ structured
modes together with the orthogonal complement:
\begin{equation}
\hat{r}_{k}=\left\langle u\,\phi\left(u\hat{h}_{k}+\sqrt{\Delta_{0}+\sum_{l\neq k}\hat{h}_{l}^{2}}z\right)\right\rangle _{DuDz}
\end{equation}
The Gaussian integral over $u$ can be performed via integration by
parts. Explicitly, one writes $w\equiv\mathrm{e}^{-\frac{u^{2}}{2}}u$
and $dv\equiv\phi\,du$, then $\int dv\,w=-\int dw\,v=\int Du\,\pdr{\phi}u$.
Now the two Gaussians, $u$ and $z$, combine to a single Gaussian
integral:
\begin{equation}
\hat{r}_{k}=\hat{h_{k}}\left\langle \phi'\left(\sqrt{\Delta_{0}+\norm{\hat{\mathbf{h}}}^{2}}x\right)\right\rangle _{Dx}\label{eq:MF structured rates}
\end{equation}
We find that due to the Gaussianity of $\mathbf{U}$, $\hat{\mathbf{h}}$
is proportional to $\hat{\mathbf{r}}$. Therefore the $D$ equations
for $\hat{h}_{k}$reduce to one implicit scalar equation for $\norm{\hat{\mathbf{h}}}$:
\begin{equation}
\norm{\hat{\mathbf{r}}}=\norm{\hat{\mathbf{h}}}\left\langle \phi'\left(\sqrt{\Delta_{0}+\norm{\hat{\mathbf{h}}}^{2}}x\right)\right\rangle _{Dx}\label{eq:Gaussian MF Subspace Firing Rate}
\end{equation}
where, of course, $\mathbf{\hat{\mathbf{r}}}$ is determined to leading
order by the balance equations.\\

The Gaussianity of $\mathbf{U}$ also simplifies the mean-field calculation
of the variance in the orthogonal complement, $\Delta_{0}$. In the
FP equation (Eqn \ref{eq:Delta Implicit Equation-1}) the sum over
neurons can be replaced by a Gaussian integral over the balance subspace
and combined with the Gaussian integral over the orthogonal complement
to yield
\begin{equation}
\Delta_{0}=g^{2}\left\langle \phi^{2}\left(\sqrt{\Delta_{0}+\norm{\hat{\mathbf{h}}}^{2}}x\right)\right\rangle _{Dx}\label{eq:Gaussian MF Delta}
\end{equation}

Similarly, the effective gain at the FP in the orthogonal complement
is given by
\begin{equation}
g_{eff}^{2}=g^{2}\left\langle \phi'^{2}\left(\sqrt{\Delta_{0}+\norm{\hat{\mathbf{h}}}^{2}}z\right)\right\rangle _{Dz}\label{eq:Gaussian MF Gain}
\end{equation}
Furthermore, we can simplify the dynamic mean-field expression for
the total autocorrelation , $C_{T}\left(\tau\right)\equiv\left\langle \phi\left(h_{i}\left(t\right)\right)\phi\left(h_{i}\left(t+\tau\right)\right)\right\rangle $,
as well. In Eqn \ref{eq:C(tau) general case}:
\begin{equation}
C_{T}\left(\tau\right)=\left\langle \left\langle \phi\left(\sqrt{\Delta_{0}-\Delta\left(\tau\right)}x+\sqrt{\Delta\left(\tau\right)+\norm{\hat{\mathbf{h}}}^{2}}y\right)\right\rangle _{Dx}^{2}\right\rangle _{Dy}
\end{equation}
Thus we find that in the Gaussian setting, the DMFT calculation of
the autocorrelation in the orthogonal subspace, $\Delta\left(\tau\right)$,
depends only on the norm in the balance subspace, $\norm{\hat{\mathbf{h}}}$.
We exploit this in order to calculate $\hat{\mathbf{C}}\left(\tau\right)$
(Eqn \ref{eq:Structure of Covariance C_hat}) as shown in Figure 4(B-D),
and we verify the calculation of $C_{T}\left(\tau\right)$ and $\Delta\left(\tau\right)$
directly in Supplementary Figure 1.\\

\subsection*{Appendix D - Constructing Connectivity Matrices with Incomplete Alignment}

\subsubsection*{D1 - Constructing Arbitrary Low-Rank Structure with Uniform Misalignment
\label{subsec:D1---Uniform Misalignment}}

As a concrete example we study a specific form of misalignment in
which the alignment matrix is scaled down uniformly across the balance
subspace modes. Concretely, for this section we fix the rank $D$
and then in order to construct the structured connectivity, $\mathbf{M}=\frac{1}{\sqrt{N}}\mathbf{U}\bb{\Sigma}\mathbf{V}^{T}$,
we first fix the diagonal of $\bb{\Sigma}$ to be non-negative numbers
(usually we set them to be all ones for simplicity), and then sample
the elements of $\mathbf{U}$ independelty from a standard Gaussian
distribution. In order to define $\mathbf{V}$, we first construct
an arbitrary orthonormal $D$-by-$D$ matrix, $\hat{\mathbf{A}}$,
which will be the alignment matrix in the case of full alignment.
To construct $\hat{\mathbf{A}}$ we generate random pairs of orthonormal
vectors of dimension $D$ which each serve as the real and imaginary
part of an eigenvector of $\hat{\mathbf{A}}$. We associate to each
pair a uniform random phase constrained to have negative real part.
For odd $D$ we add a real eigenvector with eigenvalue $-1$. Explicitly,
we generate orthonormal vectors $\mathbf{q}_{re}^{k}$ and $\mathbf{q}_{im}^{k}$
for $k=1...\left\lfloor \nicefrac{D}{2}\right\rfloor $, and $\theta_{k}\sim U\left[\frac{\pi}{2},\frac{3\pi}{2}\right]$,
and then define
\begin{equation}
\mathbf{z}^{k}=\mathbf{q}_{re}^{k}+i\mathbf{q}_{im}^{k}
\end{equation}
\begin{equation}
\mathbf{z}^{D-k}=\mathbf{q}_{re}^{k}-i\mathbf{q}_{im}^{k}
\end{equation}
And for odd $D$ we set $\mathbf{z}^{\nicefrac{(D+1)}{2}}$to be a
random real orthonormal vector and $\theta_{\nicefrac{\left(D+1\right)}{2}}=-\pi$.
Next we define $\mathbf{Z}$ to be the matrix of column vectors consisting
of $\mathbf{z}^{k}$ and $\bb{\Theta}$ to be the diagonal matrix
with $\theta_{k}$ along the diagonal, and finally we have
\[
\hat{\mathbf{A}}=\mathbf{Z}\mathrm{\mathrm{e}^{\bb{\Theta}}}\mathbf{Z}^{T}
\]
where the $^{T}$ here means conjugate-transpose. Thus $\hat{\mathbf{A}}$
is a random orthonormal $D$-by-$D$ matrix with eigenvalues constrained
to the left half of the complex plane. This matrix will define the
structure of true recurrence within the balance subspace, which will
be additionally scaled by a scalar parameter to adjust the extent
of alignment as follows.\\

We construct an $N$-by-$D$ matrix $\mathbf{A}_{\perp}$, with $\mathbf{A}_{\perp}^{T}\mathbf{A}_{\perp}=N\mathbf{I}_{DxD}$
such that $\mathbf{U}^{T}\mathbf{A}_{\perp}=\mathbf{0}$. And finally
we define $\mathbf{V}$ as:
\begin{equation}
\mathbf{V}\equiv a\mathbf{U}\hat{\mathbf{A}}+\sqrt{1-a^{2}}\mathbf{A}_{\perp}
\end{equation}

Thus the parameter $a$ scales down the alignment matrix uniformly:
\begin{equation}
\hat{\mathbf{V}}=\frac{1}{N}\mathbf{U}^{T}\mathbf{V}=a\hat{\mathbf{A}}
\end{equation}
\\
With this parameterization of the alignment matrix we have $\hat{\mathbf{r}}^{*}=-\frac{1}{a}\bb{\Sigma}^{-1}\hat{\mathbf{f}}$
for the solution to the balance equations. Plugging this into the
mean-field equation for $\norm{\hat{\mathbf{h}}^{*}}$ in the case
of Gaussian structure (Eqn \ref{eq:Gaussian MF Subspace Firing Rate},
we have:

\begin{equation}
\norm{\hat{\mathbf{h}}^{*}}\intop_{-\infty}^{\infty}Dz\,\phi'\left(\sqrt{\Delta_{0}+\norm{\hat{\mathbf{h}}^{*}}^{2}}z\right)=\frac{r_{0}}{a}\left\Vert \bb{\Sigma}^{-1}\hat{\mathbf{f}}\right\Vert 
\end{equation}
This equation is coupled with the mean-field equation for $\Delta_{0}$,
Eqn \ref{eq:Gaussian MF Delta}, and we solve this pair of equations
and then additionally calculate $g_{eff}$ (Eqn \ref{eq:Gaussian MF Gain})
and compare to simulations in Fig S1.\\

\subsubsection*{D2 - Constructing Heterogeneous Misalignment\label{subsec:D2---Heterogeneous Misalignment}}

Recall that our dynamic mean-field theory prediction for the total
temporal variance in the balance subspace is 
\begin{equation}
\left\langle \delta\hat{\mathbf{r}}^{T}\delta\hat{\mathbf{r}}\right\rangle =\trace{}\hat{\mathbf{C}}\left(0\right)=\frac{\delta C\left(0\right)}{N}\sum_{k}\left(\frac{1}{s_{k}^{2}}-1\right)
\end{equation}
as derived in Section \ref{sec:Incomplete-Align Fluctuations} of
the main text. Note that in the paramaterization of Appendix C1, all
$D$ singular values of $\hat{\mathbf{V}}$, are given by $a$. In
order to verify our theory in a more generic framework, we fix the
left- and right-singular vectors of $\hat{\mathbf{V}}$ and then construct
the $D$ singular values. In order to nevertheless restrict simulations
to a single parameter, we adjust the absolute value of the determinant
of $\hat{\mathbf{V}}$, and then set the sequence of $D$ singular
values to decay exponentially from $1$ while constraining their product.\\

Explicitly, given a fixed $\left|\det\hat{\mathbf{V}}\right|=\prod_{k=0}^{D-1}s_{k}$,
we write
\begin{equation}
l_{s}\equiv\frac{1}{D}\log\left|\det\hat{\mathbf{V}}\right|=\frac{1}{D}\sum_{k=1}^{D}\log s_{k}
\end{equation}
We then define $s_{k}=\exp\left(2k\frac{l_{s}}{\left(D-1\right)}\right)$
for $k=0,...D-1$. This indeed gives $\prod_{k=0}^{D-1}s_{k}=\exp$$\left(\frac{2\log\left|\det\hat{\mathbf{V}}\right|}{D\left(D-1\right)}\sum_{0}^{D-1}k\right)=\left|\det\hat{\mathbf{V}}\right|$.\\

For this parameterization we can write the total temporal variance
as
\begin{equation}
\left\langle \delta\hat{\mathbf{r}}^{T}\delta\hat{\mathbf{r}}\right\rangle =\frac{C\left(0\right)}{N}\left(\frac{1-\left|\det\hat{\mathbf{V}}\right|^{-\frac{4}{D-1}}}{1-\left|\det\hat{\mathbf{V}}\right|^{-\frac{4}{D\left(D-1\right)}}}-D\right)
\end{equation}
This is the theory curve displayed alongside simulation results in
Fig 4 (B-D) and Fig S2 (B-E).\\

\subsection*{Appendix E - Fluctuation Balance under Recurrent Misalignment\label{subsec:Appendix-E--Fluctuation Balance}}

We return to the balance subspace dynamics in the general setting
of incomplete misalignment (Eqn \ref{eq:Balance Subspace Dynamics})
\begin{equation}
\drdt{\hat{\mathbf{h}}}=-\hat{\mathbf{h}}+\sqrt{N}\left(\bb{\Sigma}\hat{\mathbf{V}}^{T}\hat{\mathbf{r}}+\hat{\mathbf{f}}\right)+\frac{1}{\sqrt{N}}\bb{\Sigma}\mathbf{V}_{\perp}^{T}\mathbf{r}_{\perp}\label{eq:SI Balance Subspace Dynamics}
\end{equation}
Recall that this expression for the dynamics ignores the projection
of the random connectivity onto the balance subspace, which is of
order of magnitude $O\left(\frac{1}{\sqrt{N}}\right)$.\\
\\
The dynamics of the fluctuations, $\delta\hat{\mathbf{h}}\equiv\hat{\mathbf{h}}-\left\langle \hat{\mathbf{h}}\right\rangle $,
are given by
\begin{equation}
\drdt{\delta\hat{\mathbf{h}}}=-\delta\hat{\mathbf{h}}+\bb{\Sigma}\left(\sqrt{N}\hat{\mathbf{V}}^{T}\delta\hat{\mathbf{r}}+\hat{\bb{\eta}}\right)
\end{equation}
where we have written $\hat{\bb{\eta}}\equiv\frac{1}{\sqrt{N}}\mathbf{V}_{\perp}^{T}\delta\mathbf{r}_{\perp}$

If the $O\left(1\right)$ fluctuations driven by $\hat{\bb{\eta}}$
are not canceled by corresponding fluctuations in $\delta\mathbf{r}$,
then the dynamics of $\delta\hat{\mathbf{h}}$ will have $O\left(1\right)$
fluctuations. Such significant fluctuations in the balance subspace
would drive $\hat{\mathbf{r}}\left(t\right)$ to violate the balance
equations, and would in turn generate $O\left(\sqrt{N}\right)$ fluctuations
in $\hat{\mathbf{h}}$. Therefore balance must suppress these fluctuations,
and the $O\left(1\right)$ effective feed-forward fluctuations $\hat{\bb{\eta}}$
will be canceled by fluctuations in the balance subspace activity,
$\delta\hat{\mathbf{r}}$. As shown in for example in Figure 3C in
the main text, our simulations confirm that $O\left(1\right)$ input
fluctuations from the orthogonal complement are canceled to leading
order by recurrent balance subspace input, yielding small net input
to the balance subspace.\\

This arguement yields a fluctuation balance equation:
\begin{equation}
\sqrt{N}\hat{\mathbf{V}}^{T}\delta\hat{\mathbf{r}}+\hat{\bb{\eta}}\approx0\label{eq:Fluctuation Balance-1}
\end{equation}

requiring
\begin{equation}
\delta\hat{\mathbf{r}}\approx-\frac{1}{\sqrt{N}}\left(\hat{\mathbf{V}}^{T}\right)^{-1}\hat{\bb{\eta}}
\end{equation}
to leading order.\\
\\
We employ the SVD of $\hat{\mathbf{V}}$ in order to further simplify,
writing $\hat{\mathbf{V}}=\mathbf{L}\mathbf{S}\mathbf{R}^{T}=\sum_{k=1}^{D}L_{k}s_{k}R_{k}^{T}$,
where $L_{k}$ and $R_{k}$ are the $k$th left- and right-singular
vectors of the alignment matrix, respectively. We write $\hat{\eta}_{k}^{R}\equiv R_{k}^{T}\hat{\bb{\eta}}$,
for the effective-feedforward input along the $k$th right-singular
vector of the alignment matrix, and $\delta\hat{r}_{k}^{L}\equiv L_{k}^{T}\delta\hat{\mathbf{r}}$,
for the corresponding rate fluctuations along the $k$th left-singular
vector. Then we can re-express te the fluctuation balance requirement
as
\begin{equation}
\delta\hat{r}_{k}^{L}\approx-\frac{1}{\sqrt{N}}\frac{\hat{\eta}_{k}^{R}}{s_{k}}\label{eq:SVD Basis Fluctuation Balance}
\end{equation}

Therefore, for moderate misalignment, i.e. $s_{k}\sim O\left(1\right)$,
we expect the fluctuations in the balance subspace, $\delta\hat{\mathbf{r}}$,
to be $O\left(\frac{1}{\sqrt{N}}\right)$.\\

To find $\hat{\mathbf{C}}\left(\tau\right)$ we observe that by definition
$\mathbf{V}_{\perp}$ is orthogonal to $\mathbf{U}$, and it is independent
of $\mathbf{J}$ by assumption. Therefore the fluctuations in the
effective-feedforward drive, $\hat{\bb{\eta}}$, can be approximated
as a $D$-dimensional Gaussian with matrix of cross-correlation functions:
\begin{equation}
\left\langle \hat{\bb{\eta}}\left(t\right)\hat{\bb{\eta}}^{T}\left(t+\tau\right)\right\rangle =\frac{\left(C_{T}\left(\tau\right)-C_{T}\left(\infty\right)\right)}{N}\mathbf{V}_{\perp}^{T}\mathbf{V}_{\perp}
\end{equation}
where $C_{T}\left(\tau\right)=\left\langle r_{i}\left(t\right)r_{i}\left(t+\tau\right)\right\rangle $
is the average total autocorrelation funciton of the firing activity.
For notational purposes we write $C\left(\tau\right)\equiv C_{T}\left(\tau\right)-C_{T}\left(\infty\right)=\left\langle \delta r_{i}\left(t\right)\delta r_{i}\left(t+\tau\right)\right\rangle $
in the main text (Eqn \ref{eq:Structure of Covariance C_hat}), for
$\delta r_{i}=r_{i}-\left\langle r_{i}\right\rangle $. Note that
$C\left(\tau\right)$ captures the average single-neuron temporal
variability, while the long-time autocorrelation, $C_{T}\left(\infty\right)$,
captures the spatial variabity over single-neuron average firing rates.
\\

To simplify the expression for the covariance of $\hat{\bb{\eta}}$
and derive an expression for $\hat{\mathbf{C}}\left(\tau\right)$,
we recall that $\mathbf{V}_{\perp}=\mathbf{V}-\mathbf{U}\hat{\mathbf{V}}$.
We note that the particular structure of the columns of $\mathbf{V}_{\perp}$
is unconstrained other than the requirement of being orthogonal to
every column of $\mathbf{U}$, but that the $D$-by-$D$ Gram matrix
is fully determined by $\hat{\mathbf{V}}$:
\begin{equation}
\frac{1}{N}\mathbf{V}_{\perp}^{T}\mathbf{V}_{\perp}=\mathbf{I}_{DxD}-\hat{\mathbf{V}}^{T}\hat{\mathbf{V}}=\mathbf{R}\left(\mathbf{I}_{DxD}-\mathbf{S}^{2}\right)\mathbf{R}^{T}
\end{equation}
Interestingly, this implies that the effective-feedforward drive is
uncorrelated in the basis given by the right-singular vectors of $\hat{\mathbf{V}}$,
i.e. 
\begin{equation}
\left\langle \hat{\eta}_{k}^{R}\left(t\right)\hat{\eta}_{j}^{R}\left(t+\tau\right)\right\rangle =\delta_{k,j}C\left(\tau\right)\left(1-s_{k}^{2}\right)\label{eq:FF Covariance}
\end{equation}
 We thus find, by applying Eqn \ref{eq:SVD Basis Fluctuation Balance},
that the corresponding balance subspace rate fluctuations projected
along the left-singular vectors, $L_{k}$, are uncorrelated:
\[
\left\langle \delta\hat{r}_{k}^{L}\left(t\right)\delta\hat{r}_{j}^{L}\left(t+\tau\right)\right\rangle =\delta_{k,j}\frac{C\left(\tau\right)}{N}\frac{1-s_{k}^{2}}{s_{k^{2}}}
\]
In other words, the left-singular vectors of the alignment matrix,
$L_{k}$, are the eigenvectors of the covariance, $\hat{\mathbf{C}}$,
or equivalently, the principal components of the macroscopic fluctuations
in the balance subspace. The corresponding eigenvalues, i.e. the variances,
are $\frac{C\left(\tau\right)}{N}\frac{1-s_{k}^{2}}{s_{k^{2}}}$.\\

The total variance in the balance subspace , $\trace{}\hat{\mathbf{C}}\left(0\right)=\left\langle \delta\hat{\mathbf{r}}^{T}\delta\hat{\mathbf{r}}\right\rangle $,
is 
\begin{equation}
\trace{}\hat{\mathbf{C}}\left(0\right)=\frac{C\left(0\right)}{N}\sum_{k}\left(\frac{1}{s_{k}^{2}}-1\right)\label{eq:Total Variance-1-1}
\end{equation}
\\

We can change bases to return to the standard basis of the balance
subspace (the columns of $\mathbf{U}$):\\
\begin{equation}
\hat{\mathbf{C}}\left(\tau\right)=\frac{C\left(\tau\right)}{N}\mathbf{L}\left(\mathbf{S}^{-2}-\mathbf{I}_{DxD}\right)\mathbf{L}^{T}
\end{equation}
which is equivalent to the result presented in Eqn \ref{eq:Structure of Covariance C_hat}
in the main text.\\

\subsection*{Appendix F - The Case of Non-Alignment\label{subsec:Appendix-F--NonAlignment}}

In this section we study the case of a complete non-alignment, in
which at least one of the singular values of the alignment matrix,
$\hat{\mathbf{V}}=\frac{1}{N}\mathbf{U}^{T}\mathbf{V}$, is small
($s_{j}\sim O\left(\frac{1}{\sqrt{N}}\right)$).\\

As discussed in the previous section, the Gram matrix of $\mathbf{V}_{\perp}=\mathbf{V}-\mathbf{U}\hat{\mathbf{V}}$
is determined by by $\hat{\mathbf{V}}$ and given by $\frac{1}{N}\mathbf{V}_{\perp}^{T}\mathbf{V}_{\perp}=\mathbf{R}\left(\mathbf{I}_{DxD}-\mathbf{S}^{2}\right)\mathbf{R}^{T}$.
Therefore we can find some $N$-by-$D$ matrix $\mathbf{U}_{\perp}$
whose columns have norm $\sqrt{N}$ and are all orthogonal to each
column of $\mathbf{U}$, and write
\begin{equation}
\mathbf{V}_{\perp}=\mathbf{U}_{\perp}\sqrt{\mathbf{I}_{DxD}-\mathbf{S}^{2}}\mathbf{R}^{T}
\end{equation}

Therefore we can write $\mathbf{M}=\frac{1}{\sqrt{N}}\mathbf{U}\bb{\Sigma}\mathbf{V}^{T}$
as
\begin{equation}
\mathbf{M}=\frac{1}{\sqrt{N}}\mathbf{U}\bb{\Sigma}\mathbf{R}\left(\mathbf{S}\mathbf{L}^{T}\mathbf{U}^{T}+\sqrt{\mathbf{I}_{DxD}-\mathbf{S}^{T}}\mathbf{U}_{\perp}\right)
\end{equation}

Thus, we see that in this scenario, with $s_{j}\sim O\left(\frac{1}{\sqrt{N}}\right)$,
there is one macroscopic subspace, $\mathbf{u}_{j}^{L}\equiv\mathbf{U}L_{j}$,
which does not send any recurrent feedback to the balance subspace,
and therefore as we will show, the activity in this subspace is unconstrained
by the balance equations.\\

To see this, we first rescale the balance subspace dynamics (Eqn \ref{eq:SI Balance Subspace Dynamics})
by $\bb{\Sigma}^{-1}$ and write: $\mathbf{x}\equiv\bb{\Sigma}^{-1}\hat{\mathbf{h}}$
and $\mathbf{y}\equiv\bb{\Sigma}^{-1}\hat{\mathbf{f}}$, yielding
dynamics
\[
\drdt{\mathbf{x}}=-\mathbf{x}+\sqrt{N}\left(\hat{\mathbf{V}}^{T}\hat{\mathbf{r}}+\mathbf{y}\right)
\]
to leading order, where we have momentarily ignored the effective-feedforward
input, which we will return to below.\\

Next, we rotate the rescaled dynamics to the basis of right-singular
vectors of $\hat{\mathbf{V}}$, $x_{k}^{R}\equiv R_{k}^{T}\mathbf{x}$,
while projecting the balance subspace activity to the basis of left-singular
vectors, $\hat{r}_{k}^{L}\equiv L_{k}^{T}\hat{\mathbf{r}}$, yielding
$D$ dynamical equations:

\begin{equation}
\drdt{x_{k}^{R}}\approx-x_{k}^{R}+\sqrt{N}\left(s_{k}\hat{r}_{k}^{L}+y_{k}^{R}\right)
\end{equation}
\\

Modes with non-zero alignment, $s_{k}\sim O\left(1\right)$, yield
a linear balance equation:
\begin{equation}
\hat{r}_{k}^{L}=-\frac{y_{k}^{R}}{s_{k}}\label{eq:SVD balance equations}
\end{equation}
which is a restatement of the $D$-dimensional balance equations in
the main text, using the identity $y_{k}^{R}=R_{k}^{T}\bb{\Sigma}^{-1}\hat{\mathbf{f}}$
(Eqn \ref{eq:Linear Bal Eqns}).\\
\\
For the unaligned mode, however, balance requires small external drive,
$\hat{y}_{j}^{R}\sim O\left(\frac{1}{\sqrt{N}}\right)$. That is,
we require that the strong external drive in the balance subspace
have no projection on the $j$th right-singular veector of the alignment
matrix. We note that this requirement is an extension of our assumption
throughout that $\hat{\mathbf{f}}$ is chosen in order to allow a
set of balance equations with obtainable firing rates.\\

Therefore, we can write the balance subspace firing rates as
\begin{equation}
\hat{\mathbf{r}}=-\sum_{k\neq j}\frac{y_{k}^{R}}{s_{k}}L_{k}+\hat{r}_{j}^{L}\left(t\right)L_{j}
\end{equation}

where $r_{j}^{L}\left(t\right)$ is unconstrained by the balance equations.\\

We now turn to the fluctuation dynamics of $j$th mode. We write $s_{j}=\frac{s}{\sqrt{N}}$
and $y_{j}^{R}=\frac{y}{\sqrt{N}}$, and we have
\[
\dr{x_{j}^{R}}t=-x_{j}^{R}+s\hat{r}_{j}^{L}+y+\hat{\eta}_{j}^{R}
\]
where $\hat{\eta}_{j}^{R}=\frac{1}{\sqrt{N}}R_{j}^{T}\mathbf{V}_{\perp}^{T}\mathbf{r}_{\perp}\sim O\left(1\right)$,
which in the largne $N$ limit is a Gaussian process with autocorrelation
$\left\langle \eta_{j}^{R}\left(t\right)\eta_{j}^{R}\left(t+\tau\right)\right\rangle =C_{T}\left(\tau\right)$
following Eqn \ref{eq:FF Covariance}. This equation predicts $O\left(1\right)$
fluctuations in both $x_{j}^{R}$ and $\hat{r}_{j}^{L}$, but the
self-consistent solution involves all the other modes and is beyond
the scope of this work.\\

In the limit of zero alignment, $s=y=0$, the $x_{j}^{R}$ are expected
to be simply a Gaussian process with autocorrelation given by 
\begin{equation}
\left\langle x_{j}^{R}\left(t\right)x_{j}^{R}\left(t+\tau\right)\right\rangle =\frac{\Delta\left(\tau\right)}{g^{2}}
\end{equation}
where $\Delta\left(\tau\right)\equiv\left\langle h_{i}^{\perp}\left(t\right)h_{i}^{\perp}\left(t+\tau\right)\right\rangle $.\\

Note, however, that the fluctuations in $\hat{\mathbf{h}}=\bb{\Sigma}\mathbf{x}$
will have non-trivial impact on the microscopic fluctuations in the
orthogonal subspace, and therefore $\Delta$$\left(\tau\right)$ can
no longer be derived from the DMFT theory above (Eqn \ref{eq:DMFT diff-eqn}),
except in the limit of $\sigma_{j}\ll1$. As shown in \citet{Landau2018},
in that limit the fluctuations in $\hat{h}_{j}^{R}$ do not impact
$\Delta\left(\tau\right)$ to leading order, and therefore this regime
exhibits a passive coherent chaos with macroscopic fluctuations that
are driven in a purely feedforward-like manner by the microscopic
chaos in the orthogonal complement via the non-aligned mode of the
structured connectivity.\\

\subsection*{Appendix G - Detailed Examples}

\subsubsection*{G1 - Degenerate Excitation-Inhibition Balance Example for Unequal
Population Size\label{subsec:G1---Degenerate E-I with Unequal Pop Size}}

As in the main text, we consider an E-I network in which the synaptic
weight depends only on the pre-synaptic neuron: $J_{EI}=J_{II}=J_{I}$
and $J_{EE}=J_{IE}=J_{E}$, such that the structured connectivity
becomes rank one: $\frac{\sigma}{\sqrt{N}}\mathbf{u}_{0}\mathbf{v}_{0}^{T}$.
We consider a network with $N_{I}=\gamma N$ inhibitory neurons, and
the remaining $N_{E}=\left(1-\gamma\right)N$ are excitatory. The
structured connectivity is
\begin{equation}
\mathbf{u}_{0}=\left(\begin{array}{c}
\mathbf{1}_{N_{E}}\\
\mathbf{1}_{N_{I}}
\end{array}\right)
\end{equation}
\begin{equation}
\mathbf{v}_{0}=\frac{1}{\sigma}\left(\begin{array}{c}
J_{E}\mathbf{1}_{N_{E}}\\
-J_{I}\mathbf{1}_{N_{I}}
\end{array}\right)
\end{equation}
\begin{equation}
\sigma=\sqrt{\left(1-\gamma\right)J_{E}^{2}+\gamma J_{I}^{2}}
\end{equation}
where $\mathbf{1}_{N_{x}}$ is the uniform column-vector of length
$N_{X}$.

The alignment between $\mathbf{u}_{0}$ and $\mathbf{v}_{0}$ is given
by:
\begin{equation}
\hat{v}=\frac{\left(1-\gamma\right)J_{E}-\gamma J_{I}}{\sigma}
\end{equation}
Inhibition-dominance and network stability will require that $J_{I}\ge\frac{1-\gamma}{\gamma}J_{E}$.
At the critical boundary, $\hat{v}=0$.\\

External allignment requires that the external drive be uniform, $\mathbf{f}=\mathbf{1}$,
and the balance equation yields population average firing
\begin{equation}
\hat{r}=\frac{r_{0}}{\gamma J_{I}-\left(1-\gamma\right)J_{E}}
\end{equation}
\\
The external drive can be made to compensate for diminished alignment
by scaling $r_{0}\propto\hat{v}\sigma=\gamma J_{I}-\left(1-\gamma\right)J_{E}$.
In this case even as inhibition is weakened and alignment decreases,
the balanced fixed point remains unchanged to leading-order. In this
situation, $\hat{C}\left(t\right)=\left\langle \hat{r}\left(t\right)\hat{r}\left(t+\tau\right)\right\rangle $
is given by:
\begin{equation}
\hat{C}\left(\tau\right)=\frac{C\left(\tau\right)}{N}\frac{\gamma\left(1-\gamma\right)\left(J_{E}+J_{I}\right)^{2}}{\left(\left(1-\gamma\right)J_{E}-\gamma J_{I}\right)^{2}}\label{eq:DegenEI Theory-1}
\end{equation}

\subsubsection*{G2 - Heterogeneous In- and Out-Degrees\label{subsec:G2---Heterogeneous In- and Out-Degrees}}

Here we consider the case of heterogeneity in both out- and in-degrees,
with possible correlations between them. We have a single inhibitory
population in which each neuron $i$ is randomly connected via $K_{i}^{in}$
incoming connections, and has $K_{i}^{out}$ randomly chosen outgoing
connections, with each non-zero synapse having weight $-\frac{J}{\sqrt{N}}$,
where $K$ is the average number of connections per neuron. Such a
connectivity structure can be approximated by a deterministic rank-one
structure given by
\begin{equation}
M_{ij}=-\frac{J}{\sqrt{N}}\frac{K_{i}^{in}K_{j}^{out}}{KN}
\end{equation}
We define the relative in/out degrees as $k_{i}^{\alpha}\equiv\frac{K_{i}^{\alpha}}{K}$,
and the mean-square of the relative in/out degrees is $\left\langle k_{\alpha}^{2}\right\rangle $,
for $\alpha\in\left\{ \mathrm{in},\,\mathrm{out}\right\} $. Then
we can write
\begin{equation}
\mathbf{u}=\frac{\mathbf{k}^{in}}{\sqrt{\left\langle k_{in}^{2}\right\rangle }}
\end{equation}
\begin{equation}
\mathbf{v}=-\frac{\mathbf{k}^{out}}{\sqrt{\left\langle k_{out}^{2}\right\rangle }}
\end{equation}
\begin{equation}
\sigma=\sqrt{\left\langle k_{in}^{2}\right\rangle \left\langle k_{out}^{2}\right\rangle }Jp
\end{equation}
where $p\equiv\frac{K}{N}$.\\

The scalar alignment in this case is
\begin{equation}
\hat{v}=-\frac{1}{N}\frac{\mathbf{k}^{in^{T}}\mathbf{k}^{out}}{\sqrt{\left\langle k_{in}^{2}\right\rangle \left\langle k_{out}^{2}\right\rangle }}=-\frac{1+c}{\sqrt{\left\langle k_{in}^{2}\right\rangle \left\langle k_{out}^{2}\right\rangle }}
\end{equation}
where $c=\left\langle \left(k_{i}^{in}-1\right)\left(k_{i}^{out}-1\right)\right\rangle $
is the covariance of the relative in- and out-degrees. We find that
if the in-degree and out-degrees are uncorrelated, then 
\[
\hat{v}=-\frac{1}{\sqrt{\left\langle k_{in}^{2}\right\rangle \left\langle k_{out}^{2}\right\rangle }}
\]
and the extent of alignment decreases with increasing breadth of the
degree distributions. Correlations between in- and out-degrees increase
the alignment and in the extreme case of fully correlated degrees,
the absolute alignment remains large, and depends only on the relative
breadth of the two distributions, for example $\hat{v}=-1$ for fully
correlated degree distributions with identical variances. \\
\\
External alignment requires that $\mathbf{f}=\mathbf{u}=\frac{\mathbf{k}^{in}}{\sqrt{\left\langle k_{in}^{2}\right\rangle }}$,
similar to \citep{Landau2016a}. The balance equation gives
\begin{equation}
\hat{r}=-\frac{r_{0}}{\hat{v}\sigma}=\frac{r_{0}}{\left(1+c\right)Jp}
\end{equation}
We find that if in- and out-degrees are anticorrelated the balance-rates
will be driven up.\\
\\
Note that in this setting the balance subspace is defined by the in-degrees:
$\hat{r}=\frac{1}{N}\frac{\mathbf{k}^{in}}{\sqrt{\left\langle k_{in}^{2}\right\rangle }}^{T}\mathbf{r}$.
The population average firing rate, $\bar{r}\equiv\frac{1}{N}\mathbf{1}^{T}\mathbf{r}$
will be approximately equal to $\bar{r}\approx\frac{1}{\sqrt{\left\langle k_{in}^{2}\right\rangle }}\hat{r}$
. The population average external drive is scaled down by the same
factor, $\bar{r}_{0}\equiv\frac{1}{N}\mathbf{1}^{T}\mathbf{f}r_{0}=\frac{r_{0}}{\sqrt{\left\langle k_{in}^{2}\right\rangle }}$,
so that the balance fixed point is unimpacted by the degree distributions
themselves, and only affected by the in- to out- correlations.\\
\\
The coherent fluctuations, however, will increase with broader degree
distributions even in the absence of correlations:
\begin{equation}
\hat{C}\left(\tau\right)=\frac{C\left(\tau\right)}{N}\left(\left\langle k_{in}^{2}\right\rangle \left\langle k_{out}^{2}\right\rangle -1\right)\label{eq:Outdegree Theory-1}
\end{equation}
The resulting fluctuations in the population average will be $\bar{C}\left(\tau\right)=\frac{C\left(\tau\right)}{N}\left(\left\langle k_{out}^{2}\right\rangle -\frac{1}{\left\langle k_{in}^{2}\right\rangle }\right)$,
such that they increase with the breadth of each degree distribution.\\
\\
For correlated in- and out-degrees we find:
\begin{equation}
\hat{C}\left(\tau\right)=\frac{C\left(\tau\right)}{N}\left(\frac{\left\langle k_{in}^{2}\right\rangle \left\langle k_{out}^{2}\right\rangle }{\left(1+c\right)^{2}}-1\right)
\end{equation}

So that positive correlations between in- and out-degrees decrease
shared fluctuations while negative correlations amplify them. \\

\begin{figure}
\includegraphics{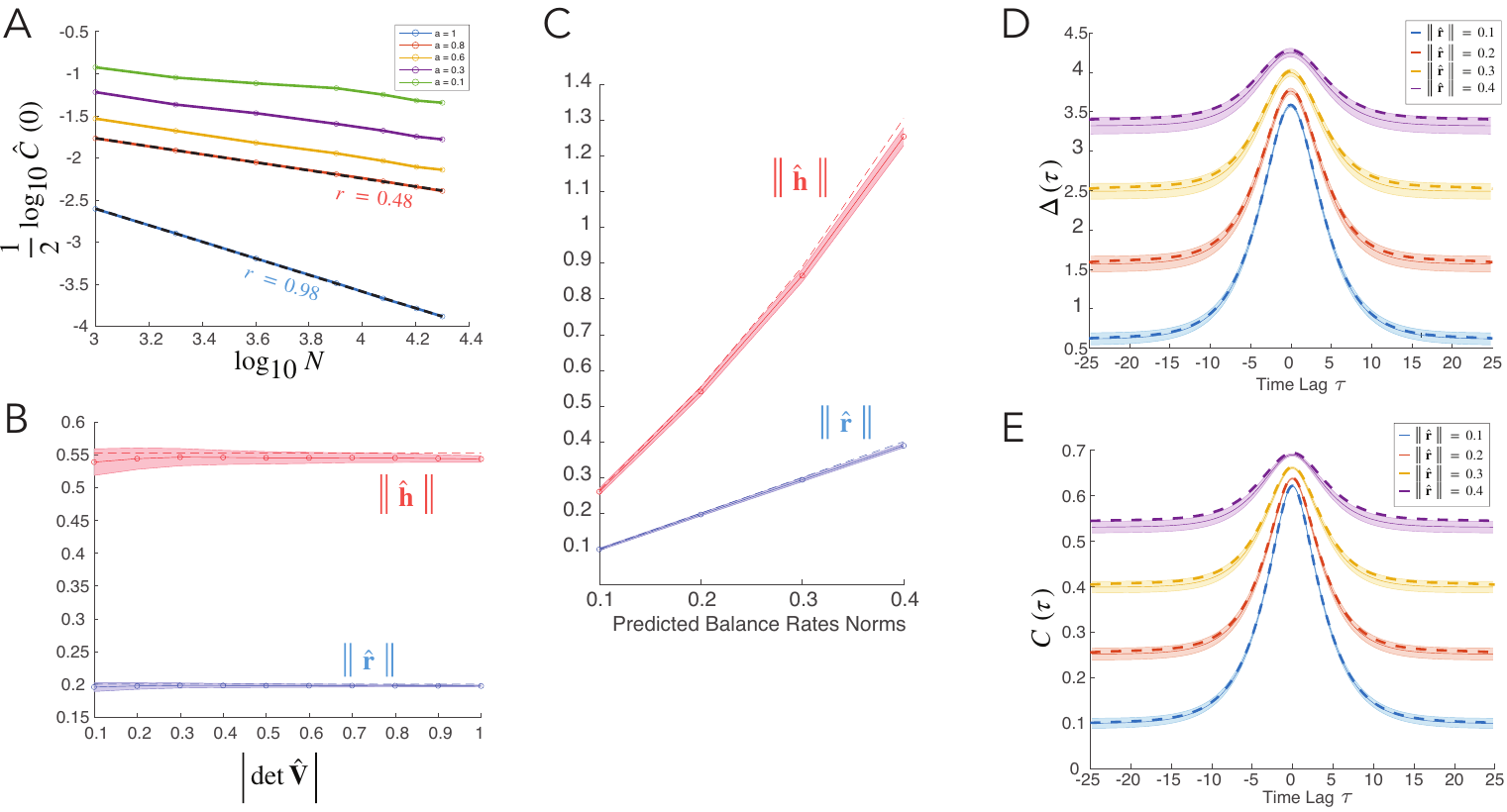}\caption{\textbf{Supplementary Figure 1: Fluctuations Due To Misalignment and
Dynamic Mean-Field Theory. (A) }Simulations of the uniform misalignment
parameterization show that misalignment of the structured connectivity
increases fluctuations in the balance subspace by an order of magnitude.
The logarithm of the standard deviation of $\hat{\mathbf{r}}\left(t\right)$
is plotted against the logarithm of network size. In the case of full
alignment ($a=1$) the slope is almost exactly $-1$, indicating that
the variance $\hat{C}\left(0\right)\sim\frac{1}{N^{2}}$. On the other
hand, even mild misalignment ($a=0.8$) yields a slope of approximately
$-0.5$, which is consistent with our theory predicting $\hat{C}\left(0\right)\sim\frac{1}{N}$
for misaligned structural connectivity. Data points are averages over
10 realizations, black dotted lines are linear fits. \textbf{(B)-(D)
}In the case of Gaussian structured connectivity, the dynamic mean-field
theory for the fluctuations in the orthogonal subspace depend only
on the norm of the firing rates in the balance subspace. \textbf{(B)
}As described in Appendix \ref{subsec:D2---Heterogeneous Misalignment},
we vary the determinant of the alignment matrix $\hat{\mathbf{V}}$
while keeping the predicted norm of the balance firing rates fixed.
This figure confirms that indeed, in our simulations both the firing
rates and the dynamical variables in the balance subspace ($\hat{\mathbf{r}}$
in blue and $\hat{\mathbf{h}}$ in red, respectively) remain constant
and very near the dynamic mean-field theory prediction throughout
the parameter range. These simulations correspond to Figure 4B. Shaded
regions display the standard deviation over 20 random realizations
of $\hat{\mathbf{V}}$ with fixed singular values, $s_{k}$ \textbf{(C)
}We vary the predicted norm of the balance firing rate while fixing
the singular values of $\hat{\mathbf{V}}$. Both $\hat{\mathbf{r}}$
in blue and $\hat{\mathbf{h}}$ in red are well-predicted by DMFT.
\textbf{(D) }The autocorrelation of the microscopic degrees of freedom
in the orthogonal complement, $\Delta\left(\tau\right)\equiv\left\langle h_{i}^{\perp}\left(t\right)h_{i}^{\perp}\left(t+\tau\right)\right\rangle $,
displayed for the same four values of $\protect\norm{\hat{\mathbf{r}}}$
as in (C), together with the corresponding DMFT prediciton. \textbf{(E)
}Same as (D) but for the autocorrelation of the singlue neuron firing
rates, $C\left(\tau\right)\equiv\left\langle r_{i}\left(t\right)r_{i}\left(t+\tau\right)\right\rangle $.
$N=10000$ accept where otherwise mentioned.}
\end{figure}
\pagebreak{}

\bibliographystyle{apsrev4-2}
\bibliography{library.bib}

\end{document}